\documentclass[twocolumn]{aastex63}

\makeatletter

\newcommand{\Rmnum}[1]{\expandafter\@slowromancap\romannumeral #1@}
\makeatother

\received{June 25, 2021}
\revised{}
\accepted{}
\submitjournal{ApJS}

\shorttitle{early-type star with emission lines}
\shortauthors{Zhang yunjin et al.}

\begin{document}

\title{A catalog of early-type H$\alpha$ emission line stars and 58 newly confirmed Herbig Ae/Bes from LAMOST DR7 }

\correspondingauthor{A-Li Luo \& Wen Hou}
\email{* lal@nao.cas.cn ~ whou@nao.cas.cn}

\author{Yun-Jun Zhang}
\affiliation{CAS Key Laboratory of Optical Astronomy, National Astronomical Observatories, Beijing 100101, China}
\affiliation{University of Chinese Academy of Sciences, Beijing 100049, China}

\author[0000-0003-0716-1029]{Wen Hou$^{*}$}
\affiliation{CAS Key Laboratory of Optical Astronomy, National Astronomical Observatories, Beijing 100101, China}

\author[0000-0001-7865-2648]{A-Li Luo$^{*}$}
\affiliation{CAS Key Laboratory of Optical Astronomy, National Astronomical Observatories, Beijing 100101, China}
\affiliation{University of Chinese Academy of Sciences, Beijing 100049, China}
\affiliation{College of computer and Information Management \& Institute for Astronomical Science, Dezhou University, Dezhou 253023, China}

\author{Shuo Li}
\affiliation{CAS Key Laboratory of Optical Astronomy, National Astronomical Observatories, Beijing 100101, China}
\affiliation{University of Chinese Academy of Sciences, Beijing 100049, China}

\author{Li Qin}
\affiliation{CAS Key Laboratory of Optical Astronomy, National Astronomical Observatories, Beijing 100101, China}
\affiliation{University of Chinese Academy of Sciences, Beijing 100049, China}
\affiliation{College of computer and Information Management \& Institute for Astronomical Science, Dezhou University, Dezhou 253023, China}

\author{Yan Lu}
\affiliation{CAS Key Laboratory of Optical Astronomy, National Astronomical Observatories, Beijing 100101, China}
\affiliation{University of Chinese Academy of Sciences, Beijing 100049, China}
\affiliation{College of computer and Information Management \& Institute for Astronomical Science, Dezhou University, Dezhou 253023, China}

\author[0000-0001-7607-2666]{Yin-Bi Li}
\affiliation{CAS Key Laboratory of Optical Astronomy, National Astronomical Observatories, Beijing 100101, China}
\affiliation{University of Chinese Academy of Sciences, Beijing 100049, China}

\author{Jian-Jun Chen}
\affiliation{CAS Key Laboratory of Optical Astronomy, National Astronomical Observatories, Beijing 100101, China}
\affiliation{University of Chinese Academy of Sciences, Beijing 100049, China}

\author{Yong-Heng Zhao}
\affiliation{CAS Key Laboratory of Optical Astronomy, National Astronomical Observatories, Beijing 100101, China}
\affiliation{University of Chinese Academy of Sciences, Beijing 100049, China}



\begin{abstract}

We derive a catalog of early-type emission-line stars including 30,048 spectra of 25,886 stars from LAMOST DR7, in which 3,922 have Simbad records. The sample is obtained using K-Nearest Neighbor and Random Forest methods and visually inspected. The spectra are classified into 3 morphological types (10 subtypes) based on H$\alpha$ emission line profiles. Some spectra contaminated by nebula emission lines such as from HII regions are flagged in the catalog. We also provide a specific sub-catalog of 101 stars with the stellar wind by calculating stellar wind or accretion flow velocities based on P Cygni or inverse P Cygni profiles, in which 74\% of them having velocities below 400km/s. More important, with two color-color diagrams (H-K, K- W1) and (H-K, J-H) of a collection of known Herbig Ae/Be stars (HAeBes) and classical Ae/Be stars (CAeBes), we propose an updated criterion to separate HAeBes from CAeBes. By the criterion, we select 201 HAeBes candidates and 5,547 CAeBes candidates from the sample. We confirmed 66 of the 201 HAeBes by using both WISE images and LAMOST spectra and present a specific HAeBe sub-catalog, in which 58 are newly identified. In addition, the WISE colors (W1-W2, W1- W3, and W1-W4) show the distribution consistency between our confirmed HAeBes and that of the known HAeBes.  Most of the 66 confirmed HAeBes locate in the lower edge of the main sequence of hot end in the HR diagram, while the distances of about 77\%  exceed 1Kpc, which enlarges the number of far distant known HAeBes. 

\end{abstract}



\section{Introduction} \label{sec:intro}

Stars with remarkable emission lines in the optical spectra are called emission-line stars. \citet{Book.2007ASSL..342.....K} classified emission-line stars into four types, namely early-type stars (Wolf-Rayet stars, Of, Oe/Be/Ae stars, Luminous Blue Variables), late-type stars (dMe stars, Mira variables, flare stars, red giants), close binaries (Algol stars, cataclysmic variables (CV), symbiotic stars) and pre-main sequence stars (PMSs, including Herbig Ae/Be stars and T Tauri stars). Emission line stars distribute widely on the Hertzsprung-Russell (HR) diagram due to various stellar types. The most common feature of emission line stars is H$\alpha$ emission line. Apart from rare stellar type stars like Wolf-Rayet stars (WRs) and Luminous Blue Variables (LBVs), the H$\alpha$ emission stars with early spectral type mainly belong to classical Ae/Be stars (CAeBes), Herbig Ae/Be stars (HAeBes) and close binaries.

\citet{Be.1976IAUS...70..453B} made the first definition of Be stars, including supergiants, rapidly rotating single stars, quasi planetary nebulae, and so on. \citet{Be.1981BeSN....4....9J} made the further definition of Be stars, usually called classical Be stars (CBes), more precisely according to luminosity class. \citet{Book.2007ASSL..342.....K} described Be stars as B-type stars showing emission lines in optical spectra, including bands of the Balmer lines, single-ionized metals and sometimes neutral helium. \citet{ReviewCBe.2003PASP..115.1153P} made a review of CBes and pointed that the term of CBes is usually used to exclude HAeBes and Algo systems. \citet{ReviewCBe.2013A&ARv..21...69R} concluded that CBes are rapidly rotating main-sequence B stars through an increasingly constrained process, forming an outwardly diffusing gaseous and dust-free Keplerian disk.

HAeBes are the early-type PMSs with intermediate-mass ($2\sim10$ $M_\odot$). \citet{Herbig.1960ApJS....4..337H} proposed a category of HAeBes for the first time and made the original definition with a list of 26 HAeBes. \citet{Herbig.1972ApJ...173..353S} made the further definition of HAeBes. \citet{1984A&AS...55..109F} provided a catalog of 57 HAeBes and candidates. Later, the sample set of HAeBes are expanded by \citet{Sample.1994ASIC..436...31T,Sample.1998A&A...331..211M,2004AJ....127.1682H}.
Infrared (IR) excess is a good criterion to separate HAeBes from CAeBes. The IR excess of CAeBes is known as the result of hot ionized gas, while the IR excess of HAeBes is generated from the thermal emission of dust. The IR excess of HAeBes is much larger than that of CAeBes (\citet{1984A&AS...55..109F}).

In recent years, there are many searches of  emission-line stars in massive datasets. \citet{2016RAA....16..138H} presented a catalog of 11,204 spectra of emission-line stars and a sublist of 23 HAeBes, 5,594 CAeBes based on LAMOST DR2. \citet{LAMOSTDR2.2020A&A...643A.122S} applied an active learning classification method to discern emission-line objects automatically based on LAMOST DR2 and provided a list of 1,013 spectra of 948 new emission-line candidates. \citet{2021MNRAS.501.5927A} identified 159 classical Ae stars based on the data from LAMOST DR5. \citet{Gaia.2020A&A...638A..21V} used artificial neural networks(ANN) and provided a catalog  based on Gaia DR2 and other sources(2MASS, WISE, and IPHAS or VPHAS+). The catalog includes 693 CBe candidates and 8,470 PMS candidates, 1,361 of which are HAeBe candidates.  \citet{GALAH.2021MNRAS.500.4849C} discovered 10,364 candidate spectra with emission components based on GALAH survey, using the method of neural network autoencoder.

Emission line stars can be divided into several types according to the morphology of emission lines. \citet{emissiontype.2010ApJS..187..228S} generated a set of synthetic Be profiles with H$\alpha$ emission lines with a non-LTE model. \citet{2016RAA....16..138H} divided the emission-line stars into six groups according to the profile of H$\alpha$ emission lines. Among various profiles of emission lines, P Cygni is a classical type charactered by an extra violet shifted absorption next to the emission line. The name comes from the supergiant P Cygni, a B type emission-line star belonging to LBV. A number of investigations about P Cygni have been made for decades(\citet{PC.1969MNRAS.144..235H,PC.1978ApJ...221..186V,PC.2002A&A...390..213M,PC.2018MNRAS.481..793M}). The P Cygni profiles usually appear in the emission lines of Balmer series, basically formed in an expanding envelope surrounding the central star. \citet{PCT.1986IAUS..116..157L} firstly defined the class of P Cygni Type(PCT) stars and pointed out the terminal velocity of wind is in the range of $100\sim400$ km/s. \citet{AtlasPCyg.1979ApJS...39..481C,SEIPCygni1987ApJ...314..726L,PN.1989ApJ...345..339C,PCygni.1997A&A...326.1117N} calculated stellar wind velocities of stars with P Cygni profiles, using profile fitting based on different physical models.

The contamination of nebula emission lines makes great trouble in searching for emission-line stars. Ionized nebulae (planetary nebulae(PNe), supernova remnants and H\Rmnum{2} regions) may contaminate the spectra in the projection of sight line with H$\alpha$, N\Rmnum{2}, S\Rmnum{2}, and O\Rmnum{3} emissions (\citet{elements.2002RMxAC..12..174R,elments.2008MNRAS.384.1045K,elements.2003A&A...400..511M}). Among kinds of ionized nebulae, H\Rmnum{2} regions are the major constituents widely spreading on Galactic disc.  As a major component of the Galactic survey, the LAMOST Spectroscopic Survey of the Galactic Anticentre (LSS-GAC) covers the region of $150^\circ \leq l \leq 210^\circ$ $|b| \leq 30^\circ$ (\citet{LAMOST.2015MNRAS.448..855Y}). A number of sources in LAMOST survey are located in the projection of H\Rmnum{2} regions, which are usually contaminated by the emission lines of H\Rmnum{2} regions. \citet{H2.2014ApJS..212....1A} provided a catalog of more than 8,000 Galactic H\Rmnum{2} regions and candidates overall Galactic longitudes within $|b| \leq 8^\circ$, based on the data of all-sky Wide-Field Infrared Survey Explorer(WISE) satellite. The ongoing versions are still updated in recent years (\citet{h2.2015ApJS..221...26A,h2.2018ApJS..234...33A}).  \citet{h2.2018PASP..130k4301W} newly identified 101 Galactic H\Rmnum{2} regions from the candidates of \cite{H2.2014ApJS..212....1A} with the data of LAMOST DR5.

In this work, we searched for a sample of early-type emission-line stars including 30,048 spectra of 25,886 stars from LAMOST DR7. We make a catalog for this sample and classified the spectra into 3 types (10 subtypes) based on the morphology of H$\alpha$ emission line profiles. We flagged the spectra contaminated by nebula emission lines in the catalog. Furthermore, we estimated velocities of stellar winds or accretion flows for 101 stars with P Cygni or inverse P Cygni profiles in the spectra. Moreover, we proposed a new method to separate HAeBes from CAeBes based on two color-color diagrams (H-K,J-H) and (H-K,K-W1) using a collection of known CAeBe and HAeBes. We applied this new method to our catalog and got 235 spectra of 201 HAeBe candidates along with 6,741 spectra of 5,547 CAeBe candidates. More important, we confirmed 66 HAeBes from 201 candidates based on spectra of LAMOST DR7 and images of WISE, 58 of which are newly identified.

The outline of this paper is as follows. We introduce the details of the sample selection in Section \ref{sec:sample}. In Section \ref{sec:analysis}, we make a morphological classification of H$\alpha$ profiles and calculate the velocities of the stellar wind of stars having P Cygni profiles. In Section \ref{sec:HAeBe}, we propose a new criterion to distinguish HAeBes from CAeBes and apply it to our catalog. In Section \ref{sec:confirmation}, we confirm 66 HAeBes based on LAMOST spectra and WISE images. In Section \ref{sec:discussion}, we make a further discussion about the HAeBes we identified. Finally, a brief summary is provided in Section \ref{sec:summary}.


\section{Sample} \label{sec:sample}



\subsection{Observational Data} \label{subsec:sample}

LAMOST is a reflecting Schmidt telescope located at the Xinglong Station of the National Astronomical Observatory, China (40$^\circ$N,105$^\circ$E), with a mean aperture of 4.3m and a field of view of 5$^\circ$. Until 2021 June, DR7 published more than 10 million low-resolution spectra (R$\sim$1800) covering 3690$-$9100$\rm\AA$.

The data we used includes $\sim $500,000 low-resolution spectra of O, B, A or EM type classified by LAMOST 1D-pipeline (\citet{LAMOST.2015RAA....15.1095L}), with radial velocities provided by LAMOST DR7. Table \ref{tab:data set} shows the composition of our data set.


\begin{deluxetable}{ccccccc}
\tablenum{1}
\tablecaption{Composition of data set\label{tab:data set}}
\tablewidth{0pt}
\tablehead{
\colhead{O} & \colhead{OB} & \colhead{B} & \colhead{B6} & \colhead{B9} & \colhead{A} & \colhead{EM}
}
\startdata
122 & 73 & 17 & 8,665 & 4,088 & 485,537 & 86
\enddata
\tablecomments{This table shows the the composition of our data set. The type information are provided by LAMOST 1D-pipeline.}
\end{deluxetable}

To build a training set for Machine Learning, we randomly chose 6,000 spectra from the work of \citet{2016RAA....16..138H}, including 3,000 with H$\alpha$ emission line as positive and 3000 without H$\alpha$ emission line as negative respectively. All the 6,000 spectra have been visually checked and 85 with bad pixel or serious noise are removed. Finally, the training set includes 2,930 positive samples and 2,985 negative samples.

\subsection{Flux normalization} \label{subsec:preprocessing}

All spectra published by LAMOST have been reduced through LAMOST 2D-pipeline with telluric absorption corrected, sky background subtracted, flux and wavelength calibrated (\citet{sky.2017RAA....17...91B}). After masking the bad pixels and removing strong lines,  we fitted the red arm of spectra with a five-degree polynomial empirically to estimate the pesudo-continuum.  We defined $f_r$ as normalized flux:
\begin{equation}
f_r = {f_i}/{f_c}
\end{equation}
$f_i$ represents flux recorded in FITS and $f_c$ represents the pesudo-continuum flux. We selected the wavelength of 6553$-$6575$\rm\AA$ centered at  6564.61$\rm\AA$ (the central vacuum wavelength of H$\alpha$ ) as H$\alpha$ characteristic band.


\subsection{Sample selection} \label{subsec:preprocessing}

\subsubsection{Selection of H$\alpha$ emission-line spectra} \label{subsubsec:spectra}

The most common method to extract features is to calculate the equivalent width (EW) or full width at half maximum(FWHM) of spectral lines. However, dealing with noise data or weak features, the detection for specific features will cause serious uncertainty, and make the best selection criterion empirically  based on features is difficult. In this case, we prefer Machine Learning methods to achieve data mining tasks in big data.

\begin{figure}
\gridline{\fig{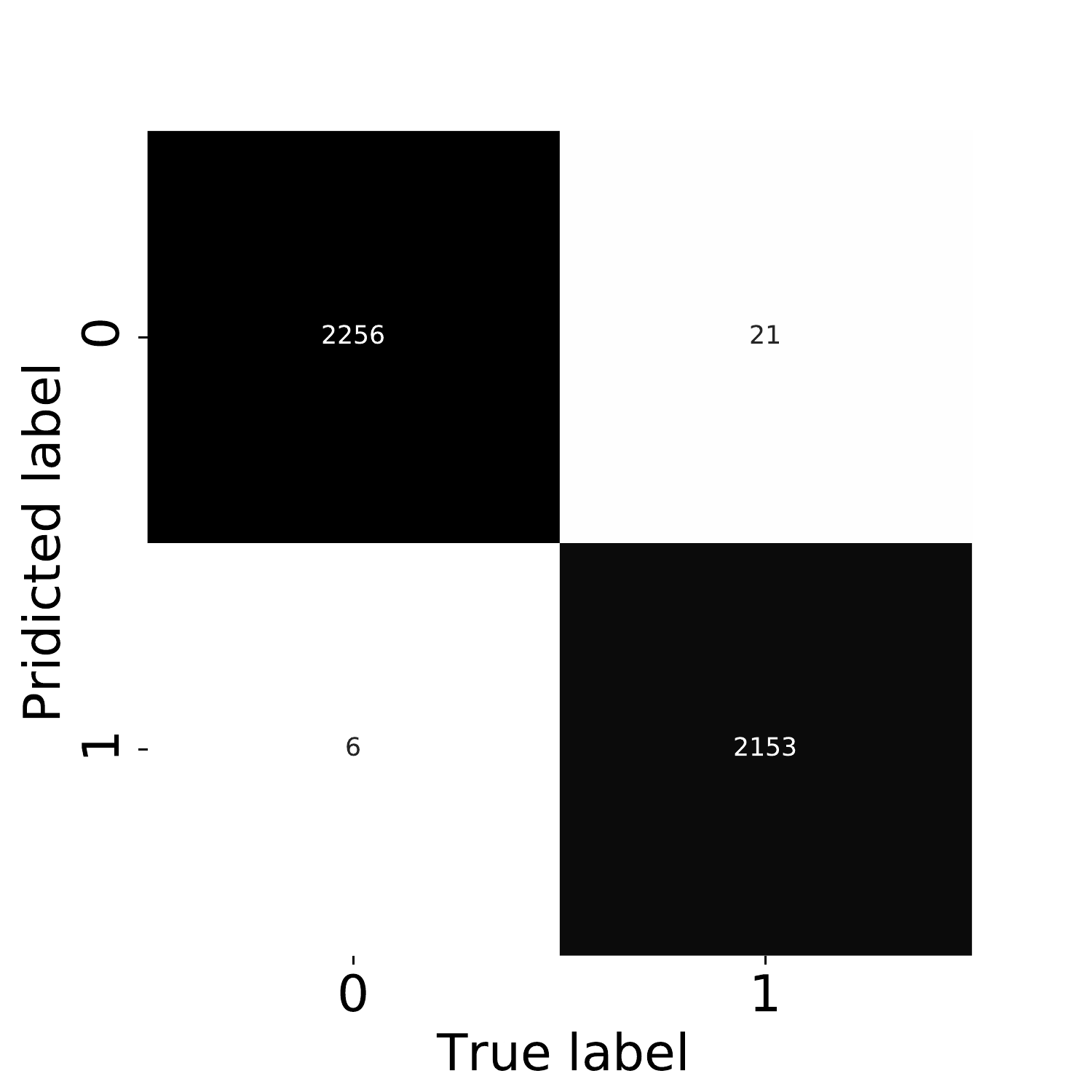}{0.25\textwidth}{Training set with KNN}
          \fig{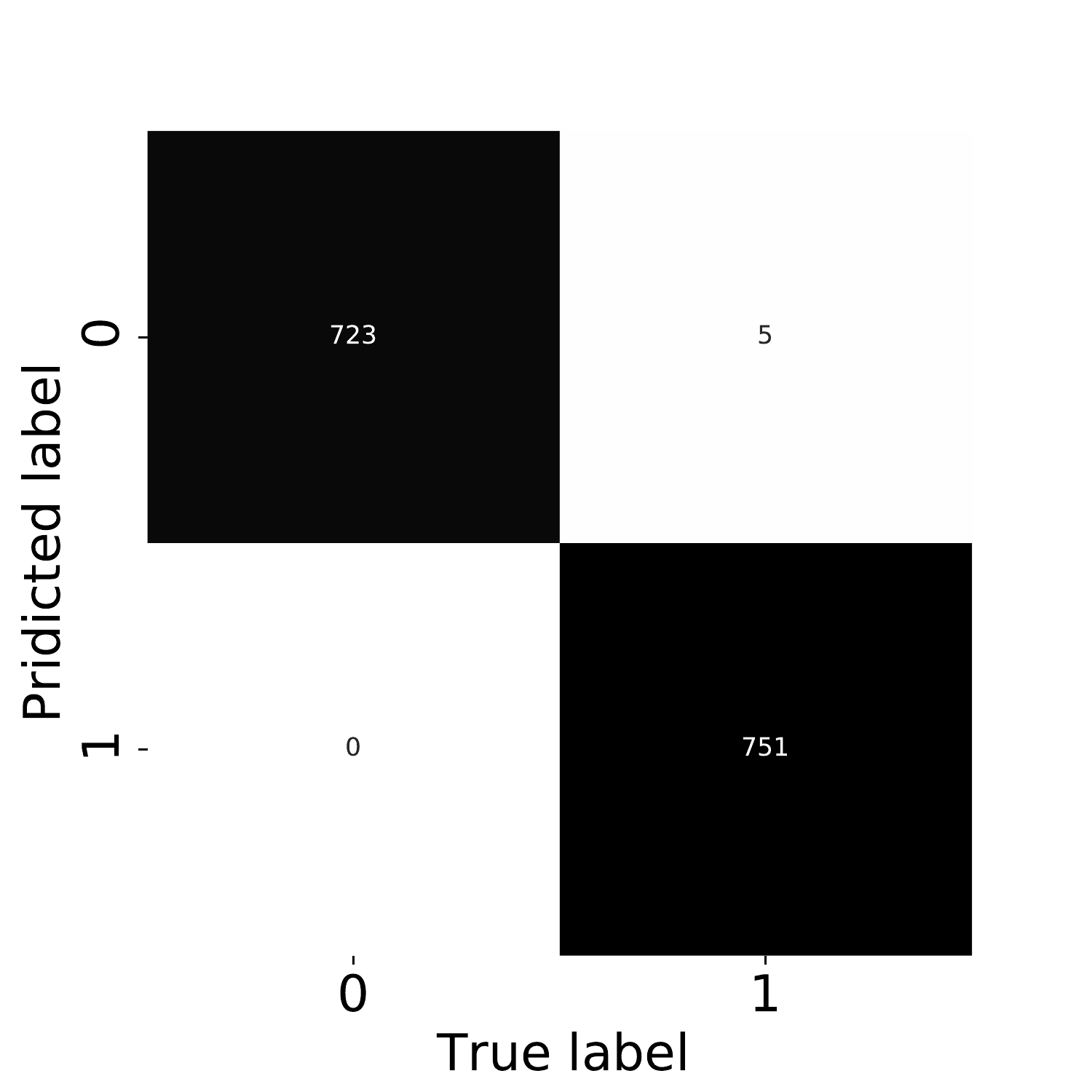}{0.25\textwidth}{Testing set with KNN}
          }
\gridline{\fig{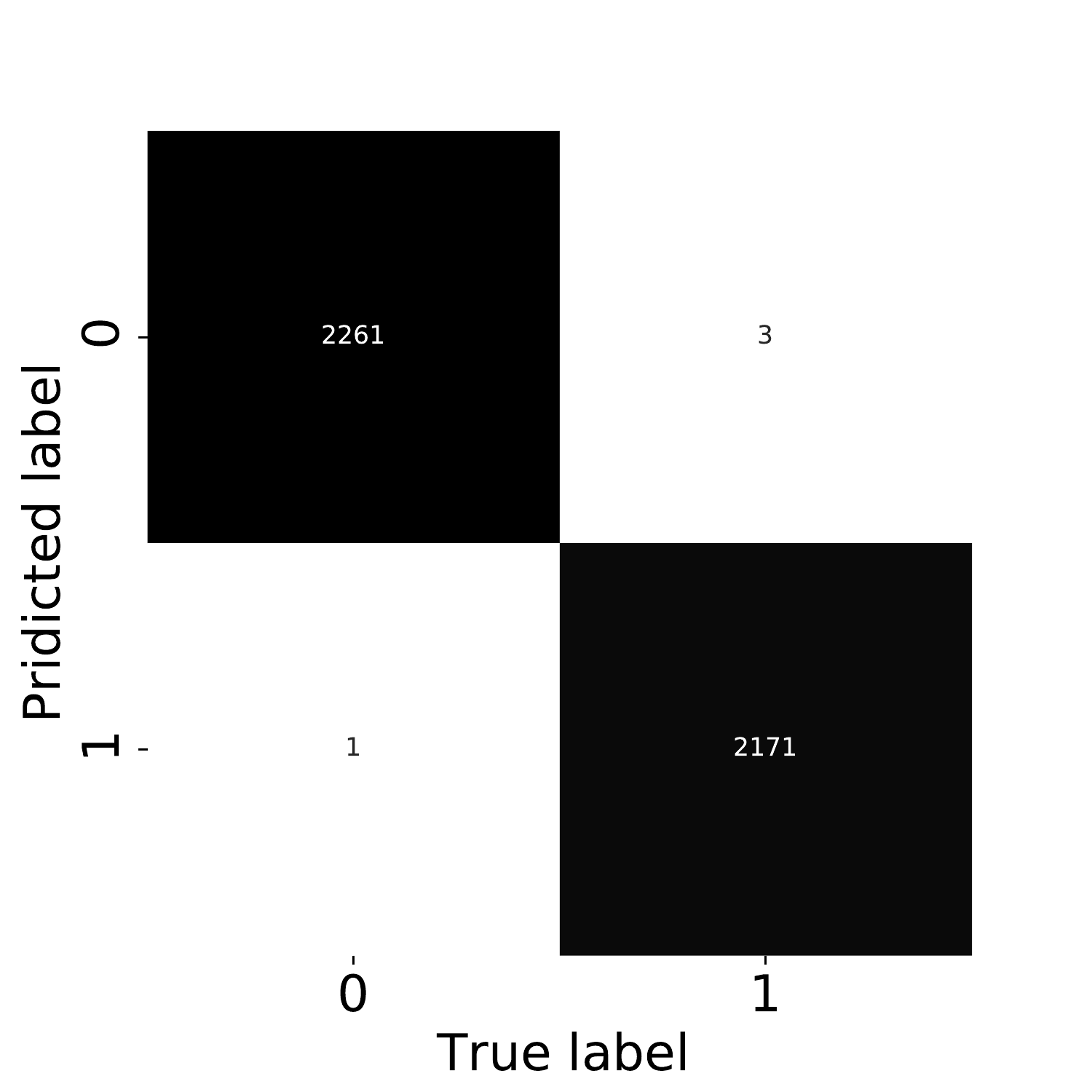}{0.25\textwidth}{Training set with RF}
          \fig{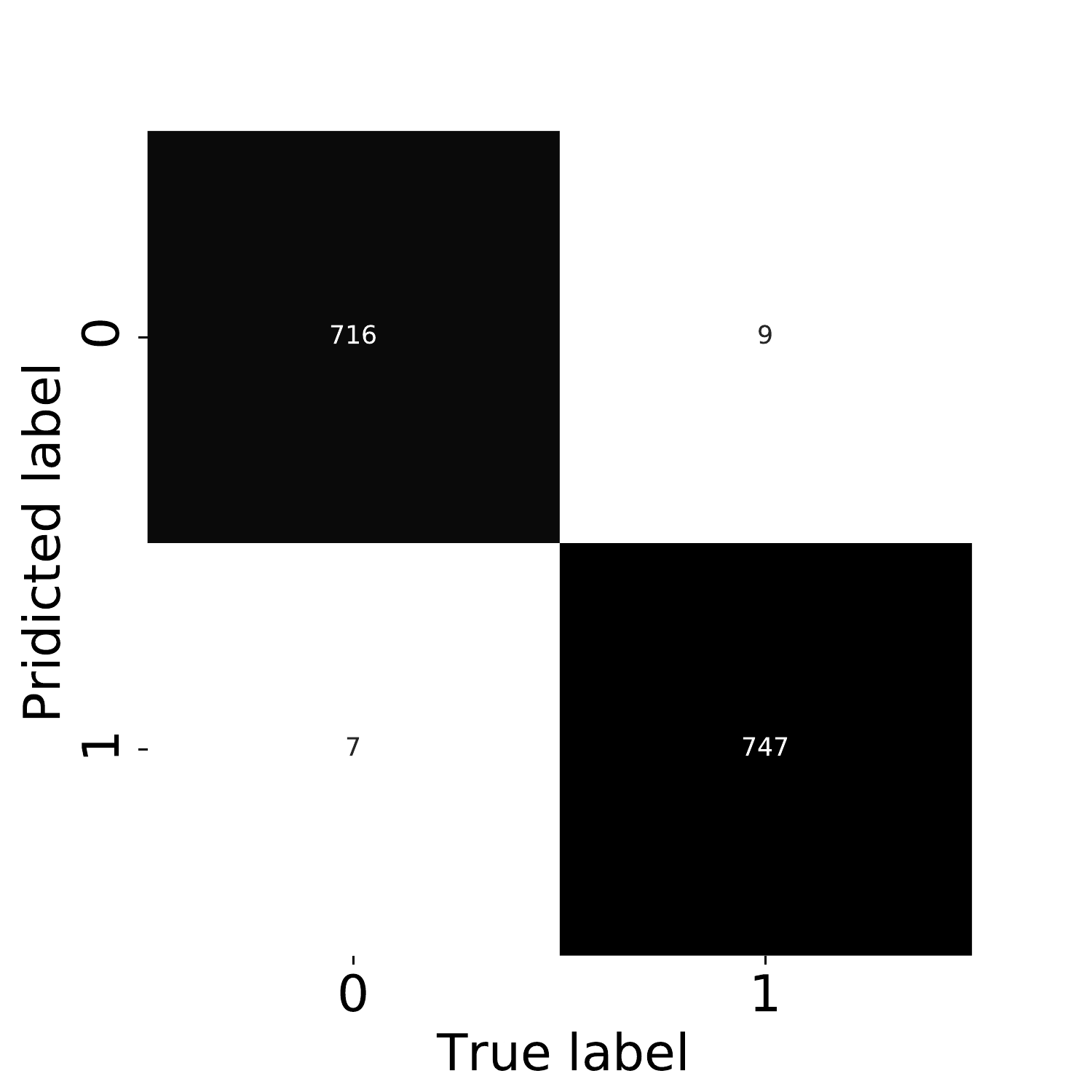}{0.25\textwidth}{Testing set with RF}
          }
\caption{Confusion matrixes of KNN (upper) and RF (bottom). The parameters of best performed KNN and RF are $K=3$, $max\_depth=15$, and $n\_estimators=1200$ respectively.
\label{fig:fig1}}
\end{figure}

We separated the 5,915 samples into a training set and testing set. Then we tested 10 machine learning methods commonly used in feature learning integrated into the software package Sklearn\footnote{\url{https://scikit-learn.org/stable/}}, including KNN (K-Nearest Neighbor), RF (Random Forest), AdaBoost, Naive Bayes (MultinomialNB, GaussianNB, BernoulliNB), logistic regression, SVM (Support Vector Machine) and Artificial Neural Network (Single-hidden Layer, Three-hidden Layer). After analysing the results, we found KNN and RF are the best among these approaches based on the combination of precision and recall.
With parametric grid searching of KNN, we adopted $K=3$ and the accuracy in the testing set is 0.997. With parametric grid searching of RF, we adopted $max\_depth=15$ and $n\_estimators=1200$, and the accuracy in the testing set is 0.989. Confusion matrixes of the training set and testing set of two methods are shown in Figure \ref{fig:fig1}.

Applying the two trained models to 498,588 spectra, we got 46,021 candidates with KNN and 54,363 candidates with RF. To include as many candidates as possible, we took the union of the two sets as our emission-line candidates, including 56,574 spectra. Table \ref{tab:S/N} lists the distribution of signal-noise-ratio (S/N) in the H$\alpha$ band of all these candidate spectra.  After visual inspection, 30,048 spectra of 25,886 sources are left.

\begin{deluxetable}{ccc}
\tablenum{2}
\tablecaption{H$\alpha$ band S/N ratio distribution of emission-line profile candidates\label{tab:S/N}}
\tablewidth{0pt}
\tablehead{
\colhead{$S/N\geq20$} & \colhead{$10<S/N<20$} & \colhead{$S/N\leq10$}
}
\startdata
26,242 & 11,360 & 18,625
\enddata
\end{deluxetable}

\subsubsection{Selection of spectra contaminated by nebula emission  lines} \label{subsubsec:nebulous}

Nebulae are interstellar clouds of dust, gas, and plasma. When a star located in nebulae, producing necessary ultra-violet photons to ionize their surrounding medium, the spectra of the surrounding medium will exhibit nebula emission  lines during de-extinction, such as H$\alpha$, N\Rmnum{2}, S\Rmnum{2}, and O\Rmnum{3}. When the target source locates in the projection of nebulae, the spectra will also be contaminated by nebula emission  lines. It's hard to determine whether the intrinsic spectra of the source have emission lines or not. We sorted and marked out spectra with nebula emission  lines as follows.

\citet{WISE.2014ApJS..212....1A} provided a catalog of H\Rmnum{2} regions based on the data of all-sky Wide-Field Infrared Survey Explorer(WISE) satellite. The catalog we referred to is the latest Version 2.2\footnote{\url{http://astro.phys.wvu.edu/wise/}}. The catalog contains 8,412 H\Rmnum{2} regions with 4 subtypes, 'K' for known, 'C' for candidate, 'G' for group, and 'Q' for radio-quiet. We cross-matched our candidates with the  H\Rmnum{2} regions catalog using galactic latitudes and longitudes. In total, 12.03\%(3,615/30,050) spectra of emission-line stars locate in the projection of H\Rmnum{2} regions, and marked as HII contamination using the labels of  \cite{WISE.2014ApJS..212....1A} in our catalog. 
When inspecting these spectra, we found most of them share a common feature of nebula emission  lines while a small fraction does not have any nebula emission  line feature which may be caused by sky subtraction during data reduction.


The above H\Rmnum{2} regions catalog mainly focus on Galactic plane, $|b|\leq8^\circ$. Besides, some other H\Rmnum{2} regions out of the catalog or regions infected by other kinds of molecular clouds may also generate nebula emission  lines overlaying on stellar spectra. In fact, this assumption has been proved by our data set. When turned to spectra not located in the H\Rmnum{2} regions provided, many spectra also show nebula emission  lines like counterpart in H\Rmnum{2} regions catalog. To select these spectra out, we measured the normalized flux of selected wavelength bands (N\Rmnum{2} 6547$-$6553$\rm\AA$, N\Rmnum{2} 6582$-$6588$\rm\AA$, S\Rmnum{2} 6715$-$6721$\rm\AA$, and S\Rmnum{2} 6730$-$6736$\rm\AA$) and detected nebula emission  lines with a visual inspection. Finally, 10,699 spectra are marked as self-recognized contamination by nebula in our catalog.

\section{Analysis of H$\alpha$ emission-line profiles} \label{sec:analysis}

\subsection{Morphological classification of H$\alpha$ profiles} \label{subsec:morphological analysis}

The morphological characteristics of H$\alpha$ profiles vary greatly in our catalog. We classified these emission-line profiles into several classes depends on morphological characteristics of H$\alpha$ emission. With the help of RF used in subclass pre-classification, 22,238 spectra of 18,965 emission-line stars have been divided into 10 subclasses of 3 main classes along with 7,810 probable Ae/Be spectra of 6,921 emission-line star candidates. The 3 main classes are Single-peak type, Double-peak type, and P Cygni or inverse P Cygni type, and the proportion of them are 92.8\%, 6.7\%, and 4.6\%, respectively. The reason of such a high proportion of Single-peak might be caused by the low resolution of our spectra.

\begin{enumerate}
\item Single-peak(Type1) is consisted of three subtypes(Type1.1, Type1.2 and Type1.3). Single-peak emission line indicates that the star is nearly pole-on i.e. the inclination is near to zero.

\begin{enumerate}
\item Without absorption (Type1.1). This type of emission-line profiles are the most explicit among 10 subclasses. The H$\alpha$ emission lines are quite remarkable without absorption profile.

\item Medium absorption (Type1.2). This type of emission-line profiles are also explicit. Dislike Type 1.1, this type has a wide shallow absorption component overlaying on the emission-line profile, and requires that the peak of emission line is above the pesudo-continuum line.

\item Deep absorption (Type1.3). This type of profiles are different from the above. An obvious emission peak located in the center of deep absorption and the peak of emission line is still under pesudo-continuum line as contrasted with Type 1.2.
\end{enumerate}

\item Double-peak (Type 2) includes five subtypes (Type 2.1, Type 2.2, Type 2.3, Type 2.4, Type 2.5). Double-peak emission line indicates that the star is mainly edge-on or the systems have high inclination.
\begin{enumerate}

\item Double-peak above continuum (Type 2.1, and 2.2). Two peaks are both above pesudo-continuna. The difference between 2.1 and 2.2 is the obviousness of double peak. 

\item Double peak below continuum (Type 2.3): Similar to 2.1 and 2.2 except that the double peaks of this type are below pesudo-continuua. 

\item Strong double-peak with deep absorption (Type 2.4, and 2.5): The type of 2.4 and 2.5 are  just like the former Type 2.1 and 2.3 respectively except that 2.4 and 2.5 have sharp and deep absorption line between two peaks.  The central star is typically edge-on. 



\end{enumerate}


\item Type 3 includes two subtypes, P Cygni profiles (Type 3.1) and inverse P Cygni profiles (Type3.2). There is an absorption component on either side of the H$\alpha$ emission line. If the absorption is on the left (Type 3.1), the blue-shifted absorption component means the envelope surrounding the central star is expanding propelled by the possible stellar wind. If the absorption is on the right (Type 3.2), the red-shifted absorption component means the envelope surrounding is contracting may be caused by gravity prompt accretion progress.
\end{enumerate}

\begin{figure*}
\gridline{\fig{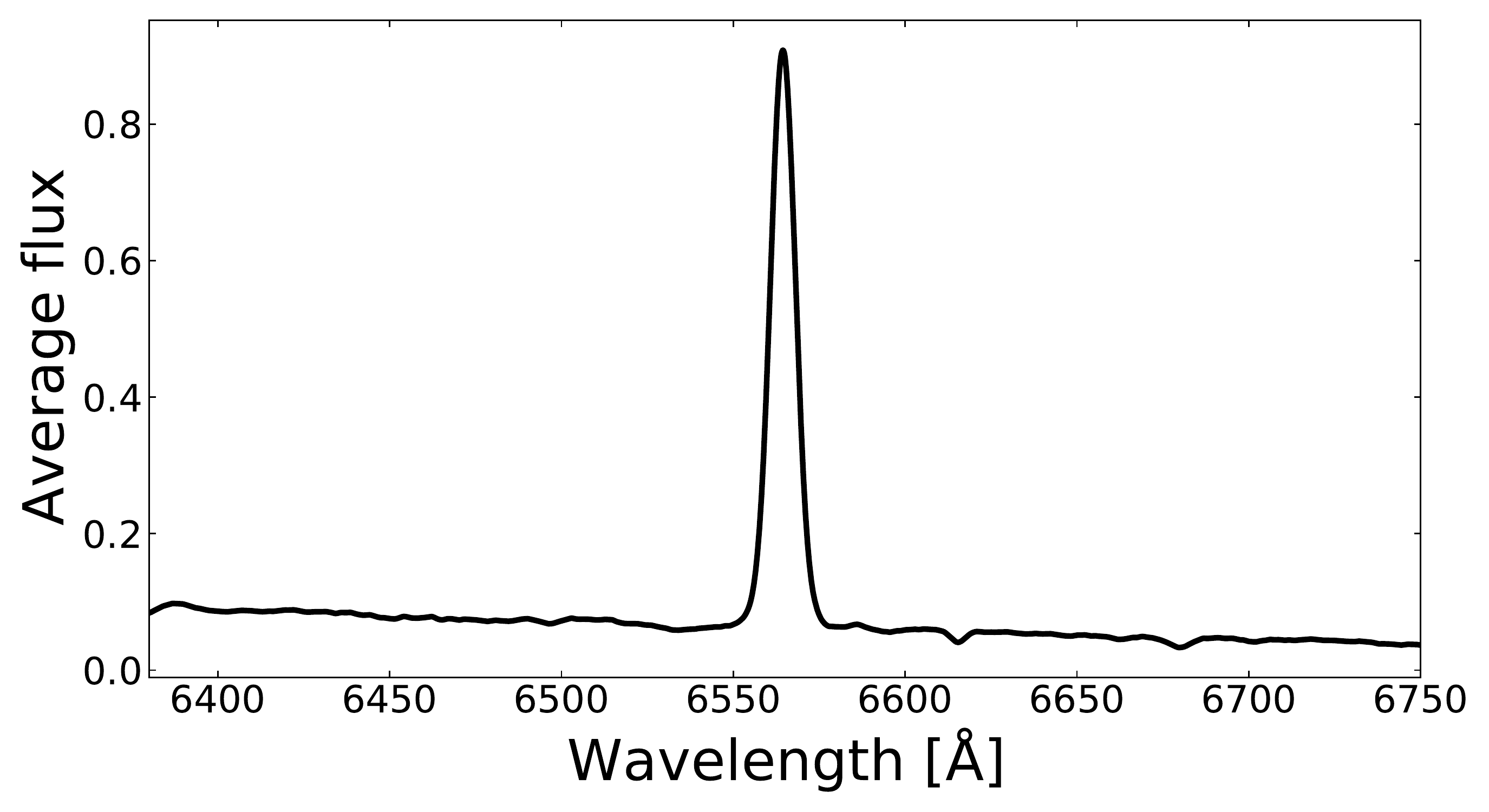}{0.333\textwidth}{Type 1.1}
          \fig{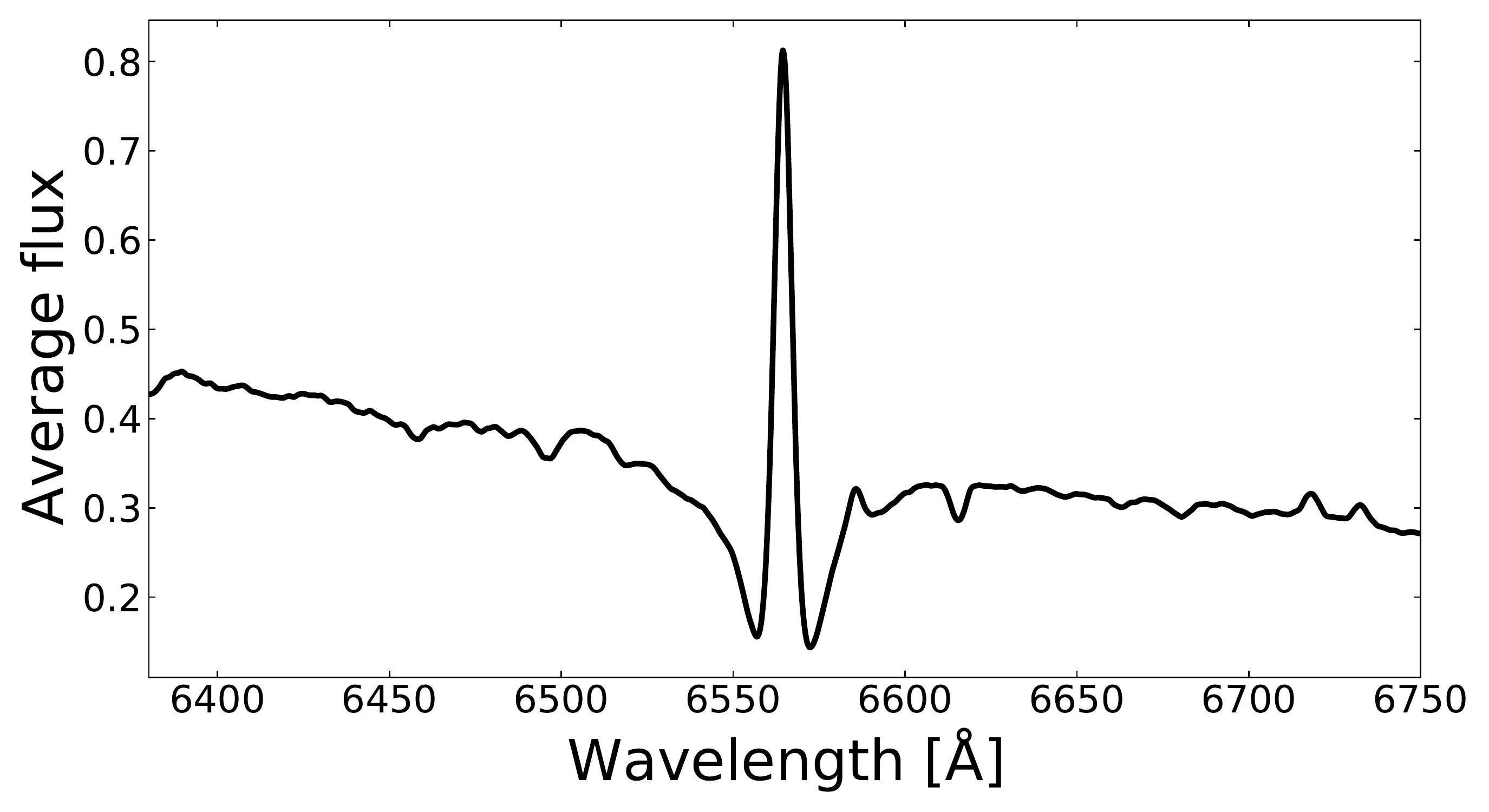}{0.333\textwidth}{Type 1.2}
          \fig{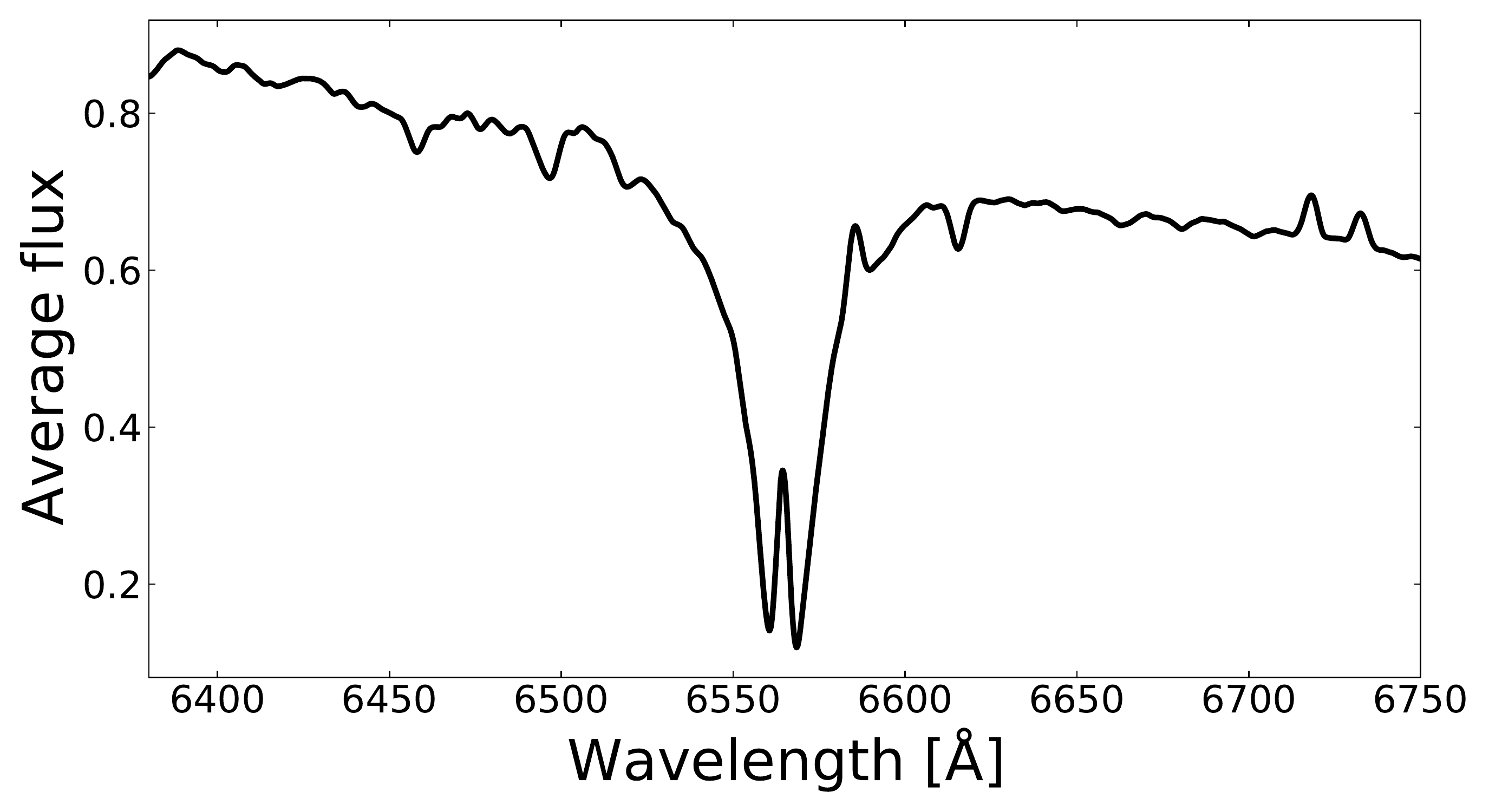}{0.333\textwidth}{Type 1.3}
          }
\gridline{\fig{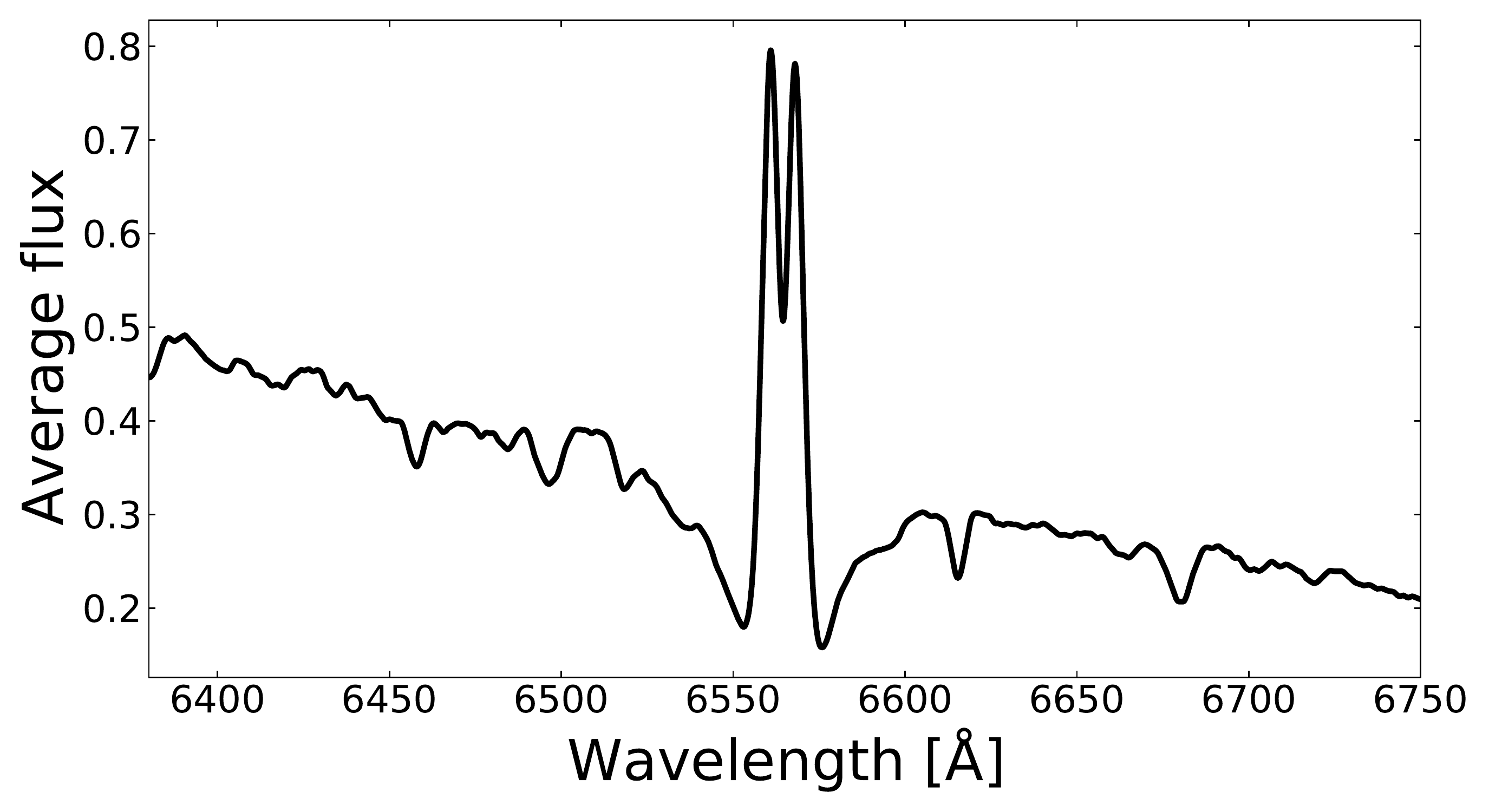}{0.333\textwidth}{Type 2.1}
           \fig{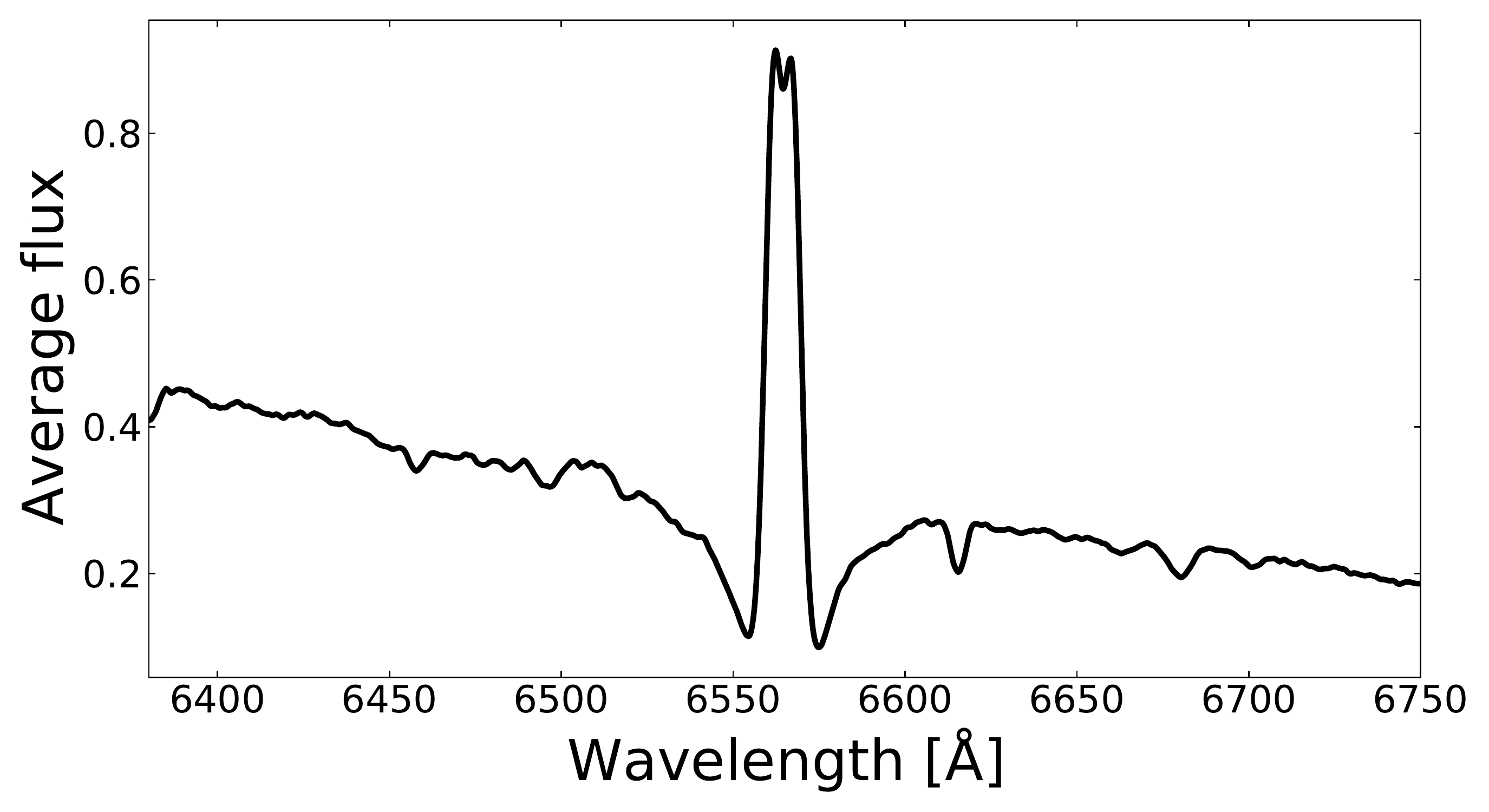}{0.333\textwidth}{Type 2.2}
           \fig{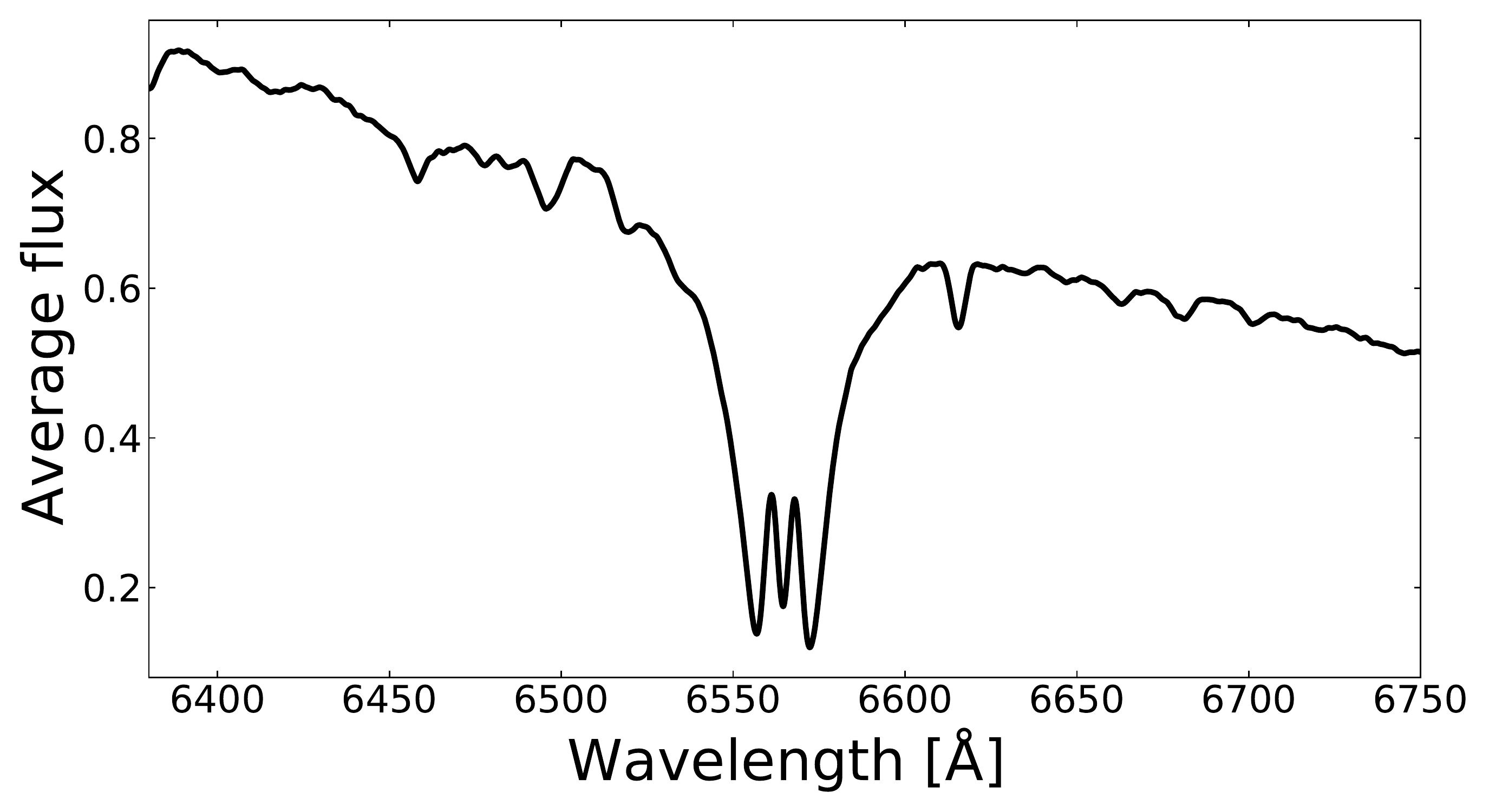}{0.333\textwidth}{Type 2.3}
          }
\gridline{
          \fig{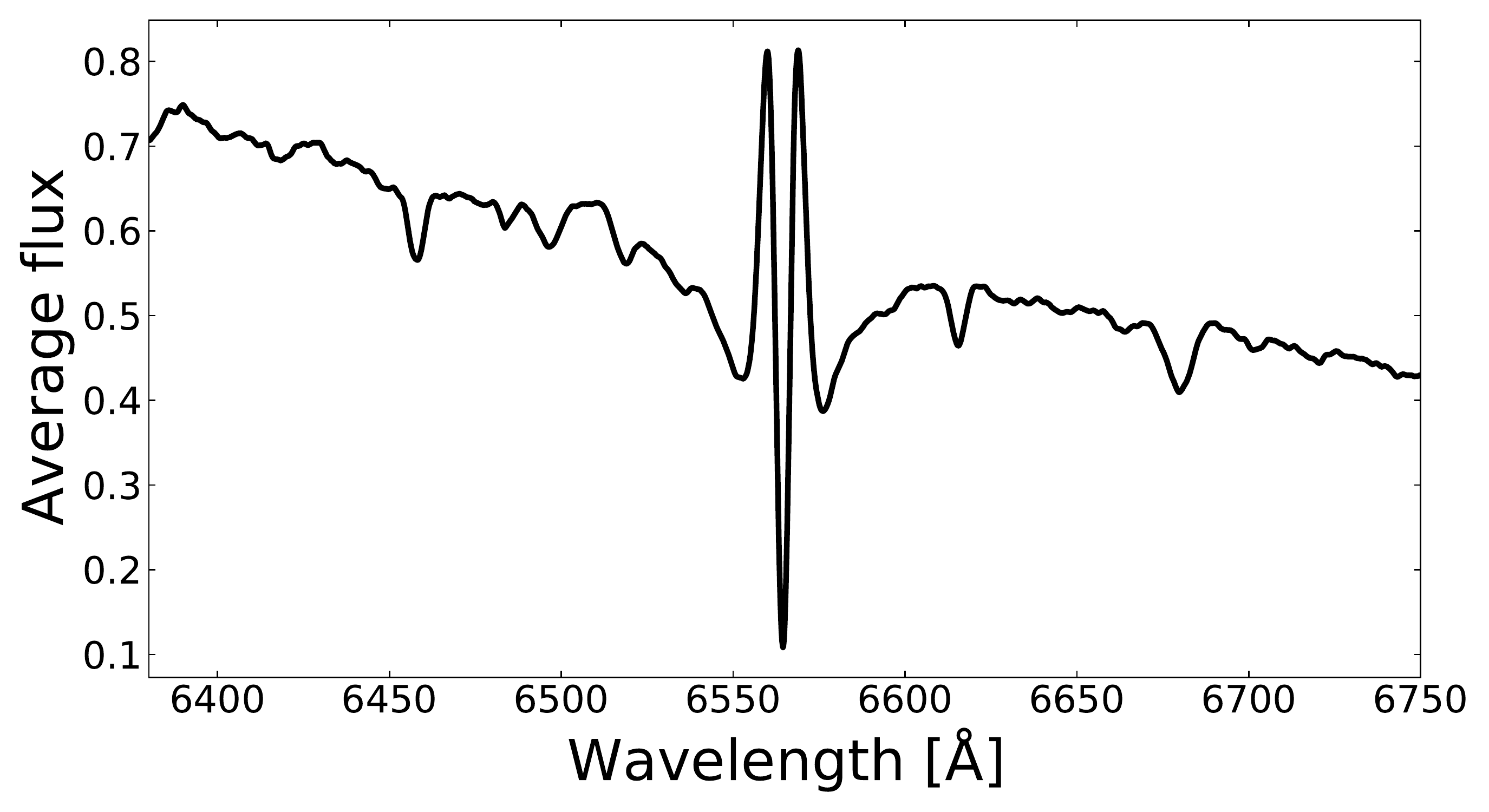}{0.333\textwidth}{Type 2.4}
          \fig{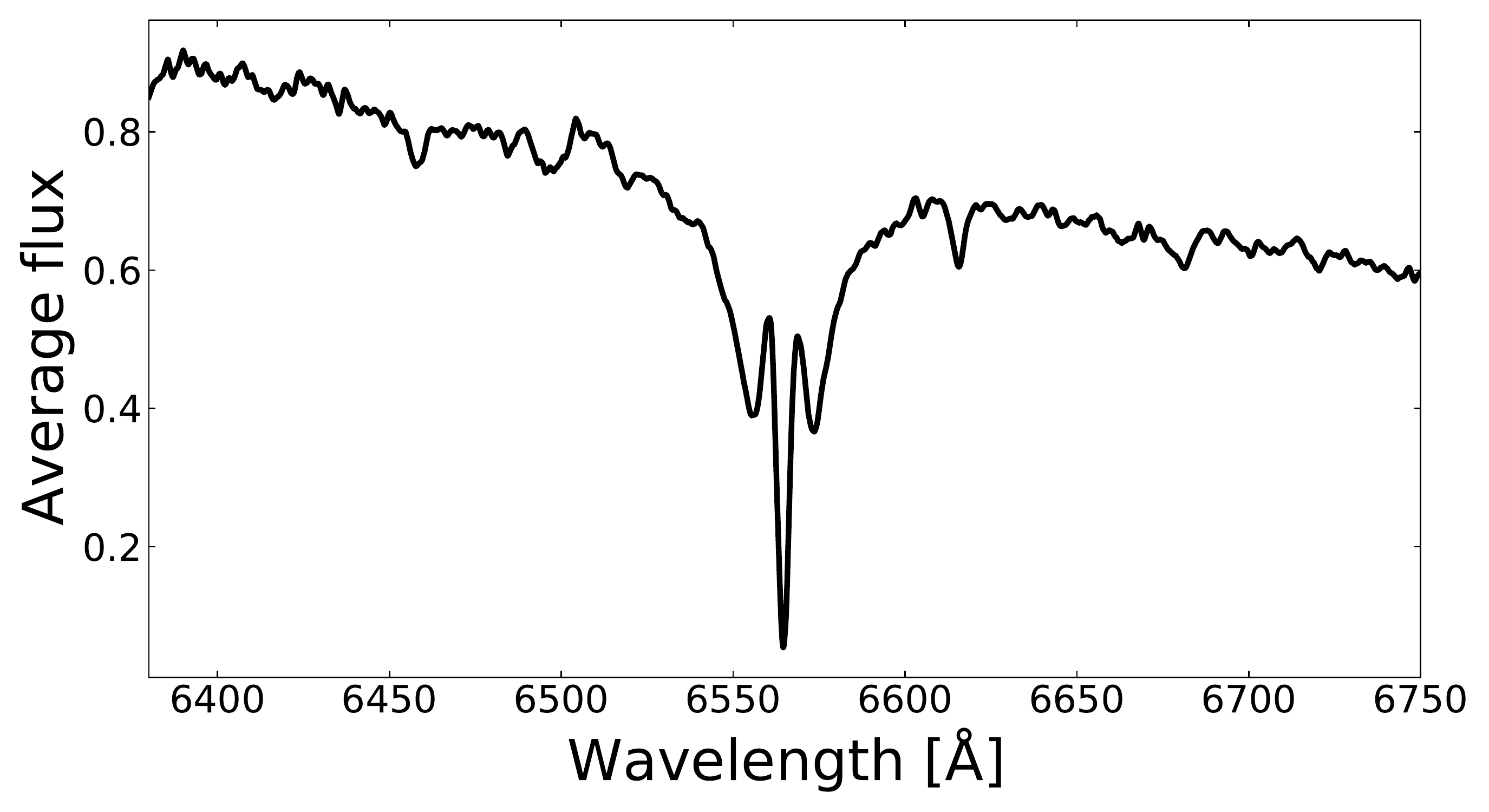}{0.333\textwidth}{Type 2.5}
          }
\gridline{\fig{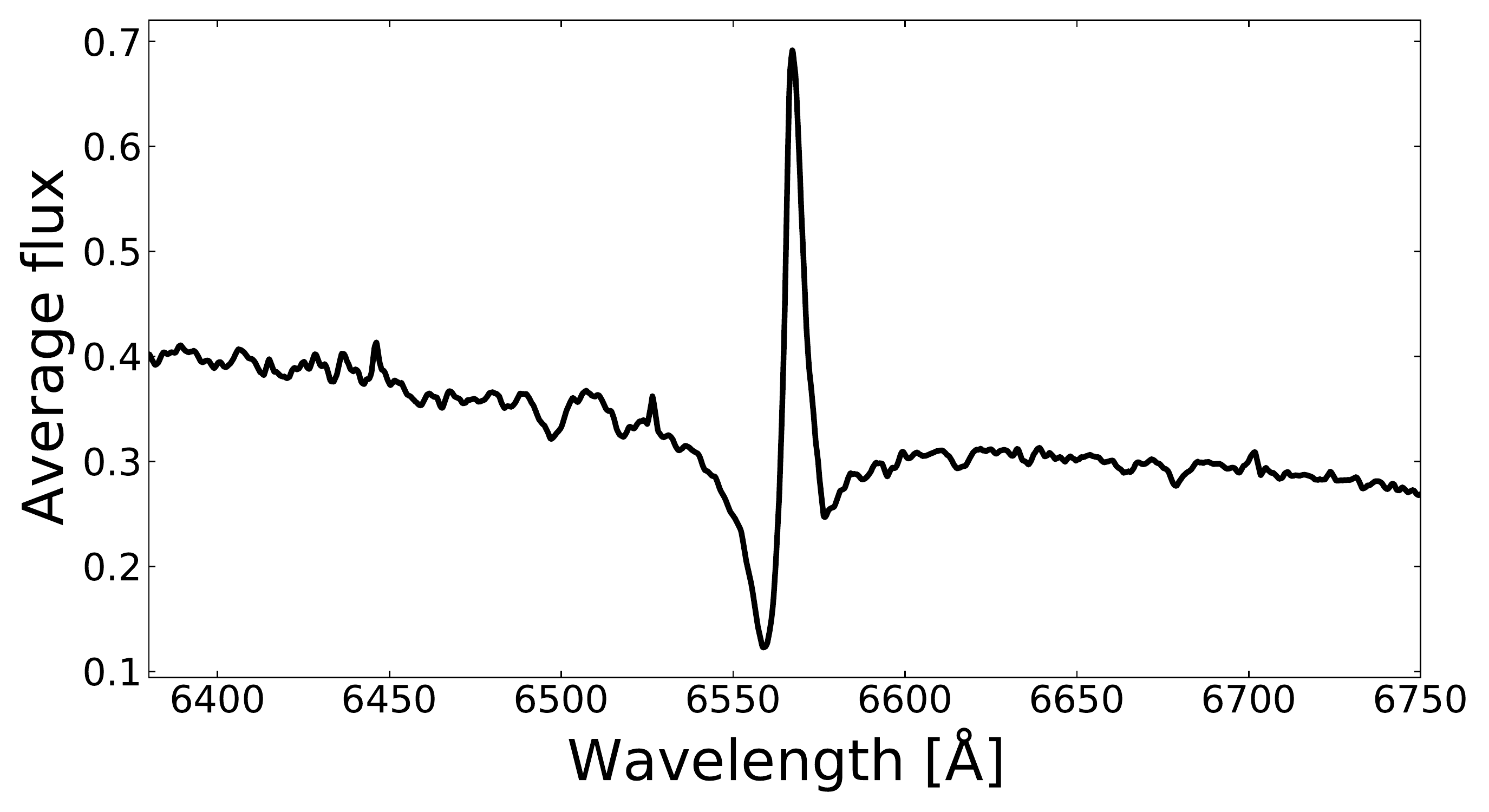}{0.333\textwidth}{Type 3.1}
          \fig{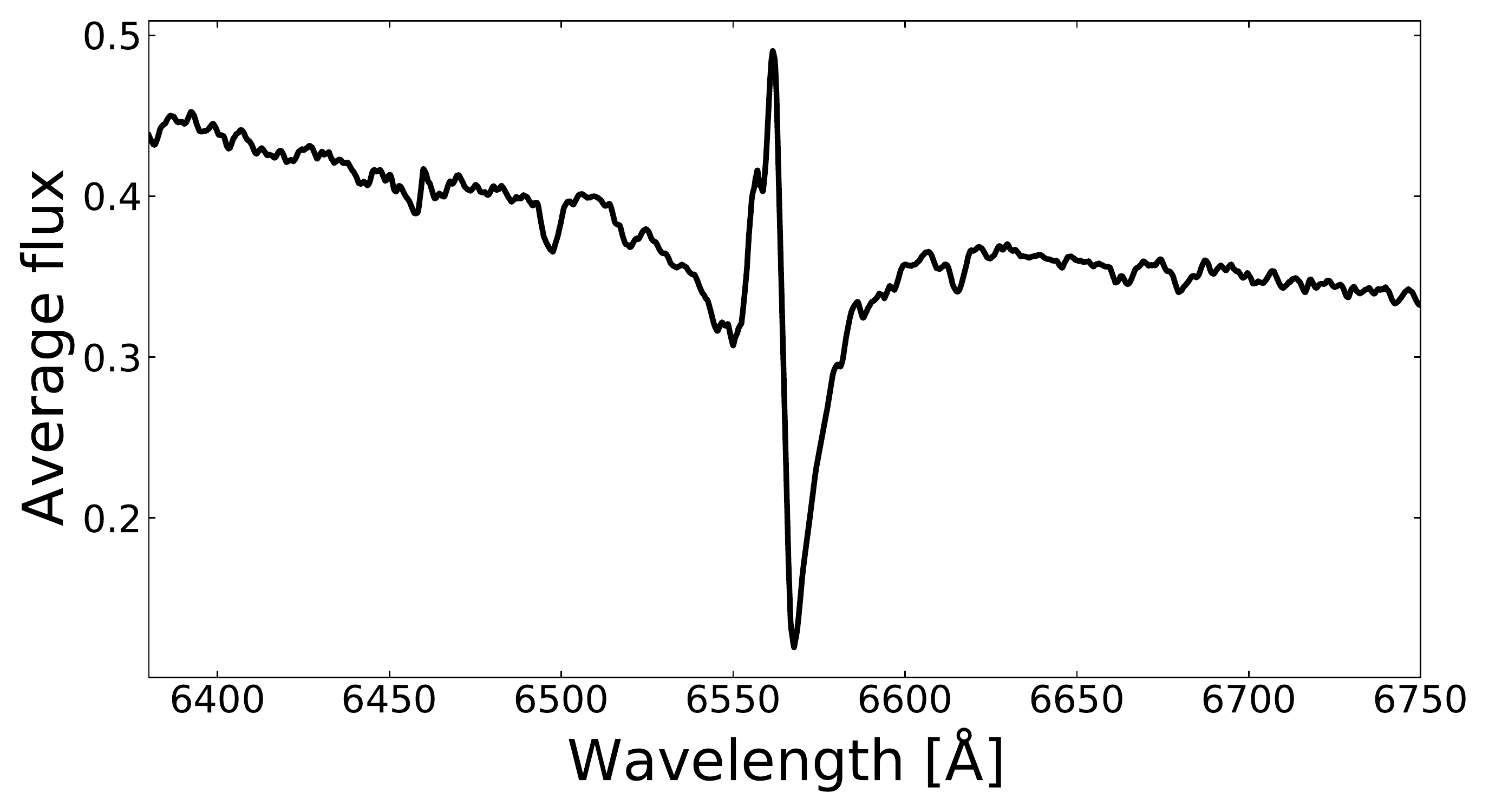}{0.333\textwidth}{Type 3.2}
          }
\caption{Mean-flux spectra of 10 subclasses based on H$\alpha$ emission profiles. Flux of each spectra have been scaled as $f_{scaled} = (f_i - f_{min})/(f_{max} - f_{min})$. $f_i$, $f_{max}$, and $f_{min}$ represent each point flux, the maximum flux, and the minimum flux of 6380$-$6750$\rm\AA$ separately.
\label{fig:fig2}}
\end{figure*}

We plotted the mean-flux spectra on H$\alpha$ band for each 10 subclasses excluding the spectra contaminated by nebula emission  lines (Section \ref{sec:sample} for details) as shown in Figure \ref{fig:fig2}.

\subsection{P Cygni profiles and stellar wind}  \label{subsec:P Cygni profile}

\subsubsection{Subsample with P Cygni profiles} \label{subsubsec:subsample}
Most spectra in the P Cygni subsample has three  H$\alpha$ components including blue-(or red-) shifted absorption,  emission, and wide absorption components. The former two are from the different regions of the disk while the latter one is from the central star. The emission line components and wide absorption lines always share the same central wavelength. We selected obvious emission with blue-(red-) shifted  absorption aside as Type 3 (P Cygni profiles as type 3.1, inverse P Cygni profiles as type 3.2). Finally, we got 103 spectra of 101 sources and estimated stellar wind or accretion flow velocities of them, seen in Table \ref{tab:velocities}. In this class, spectra with P Cygni profiles account for 40\% while those with inverse P Cygni profiles account for 60\%.



\begin{deluxetable*}{ccccccc}
\tablenum{3}
\tablecaption{Stellar wind or accretion flow velocities estimated by spectra with P Cygni or inverse P Cygni profiles\label{tab:velocities}}
\tablewidth{0pt}
\tablehead{
\colhead{ } & \colhead{ } &  \colhead{ra} & \colhead{dec} & \colhead{velocity} & \colhead{ } & \colhead{ }\\
\colhead{Designation} & \colhead{Simbad\_ID} & \colhead{[deg]} & \colhead{[deg]} & \colhead{[km/s]} & \colhead{H$\alpha$ type} & \colhead{LAMOST\_class}
}
\startdata
J012529.59+213630.6 & PG 0122+214 & 21.37332 & 21.608517 & 364.9 & 3.1 & B9\\
J013403.08+365813.8 &   & 23.512871 & 36.970506 & 405.2 & 3.1 & A7V\\
J014755.49+494802.1 &   & 26.981232 & 49.800606 & -212 & 3.2 & A7V\\
J021547.65+441943.3 &   & 33.948563 & 44.328719 & 189.2 & 3.1 & A1V\\
J031607.80+553358.0 &   & 49.032528 & 55.56613 & -212.5 & 3.2 & B6\\
J032624.73+353310.0 & HD 278643 & 51.603061 & 35.552788 & 423.9 & 3.1 & A1V\\
J034746.90+475222.1 &   & 56.945424 & 47.872812 & -234 & 3.2 & A2V\\
J034753.05+291200.0 & TYC 1812-1031-1 & 56.971046 & 29.200009 & 393 & 3.1 & A1V\\
                    &                 &           &           & 398.8 & 3.1 & A1V\\
                    &                 &           &           & 378.5 & 3.1 & A1V\\
J035426.11+480858.8 &   & 58.608812 & 48.149693 & -428 & 3.2 & A7V
\enddata
\tablecomments{The 'Simbad\_ID' column is the source name retrieved from the Simbad. The 'LAMOST\_class' column is the result of LAMOST 1D-pipeline. Complete list is available in electronic form.}
\end{deluxetable*}

\subsubsection{Velocities of stellar winds or accretion flows\label{subsubsec:velocities}}

\begin{figure*}[ht!]

\gridline{\fig{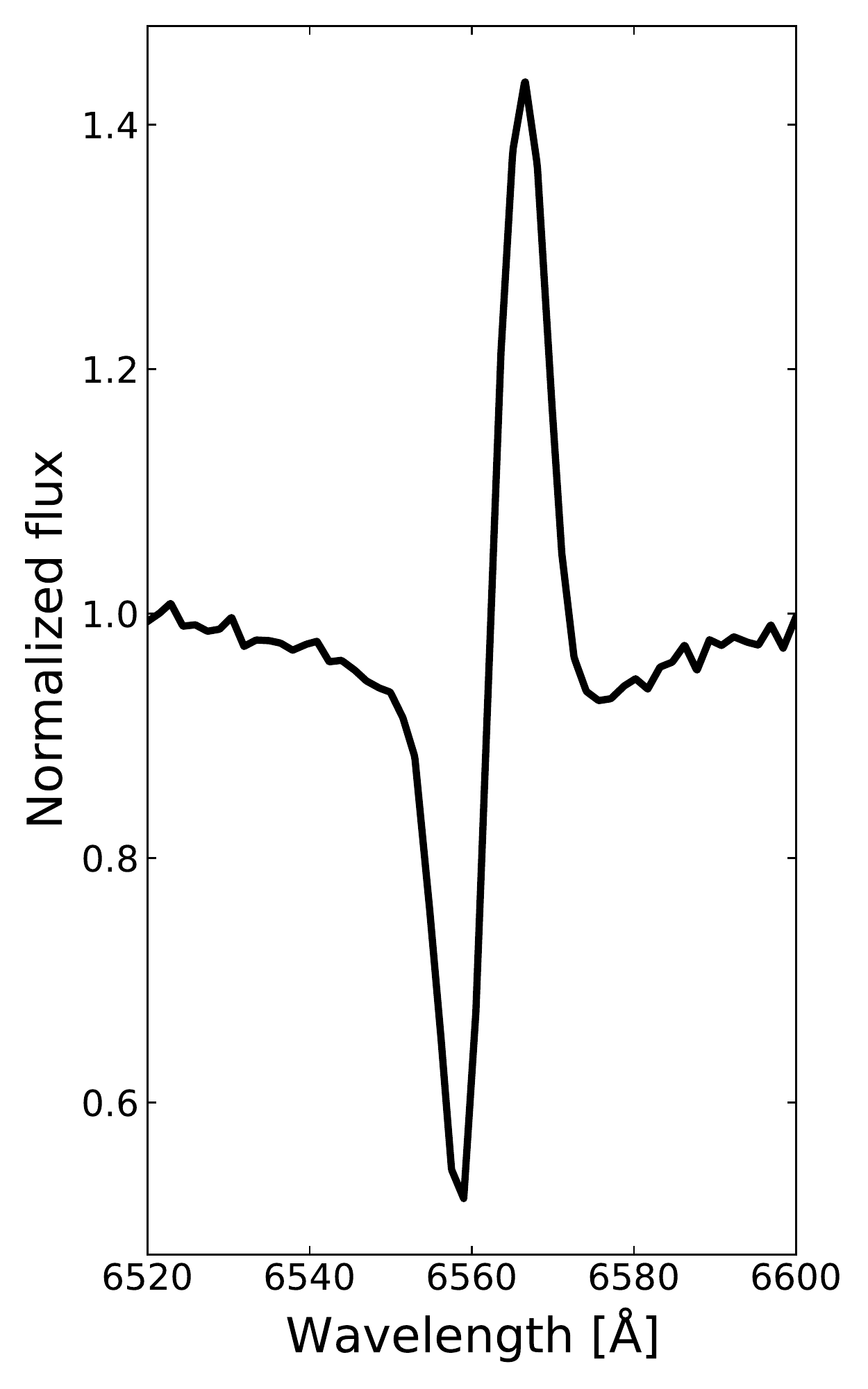}{0.2\textwidth}{Original H$\alpha$}
          \fig{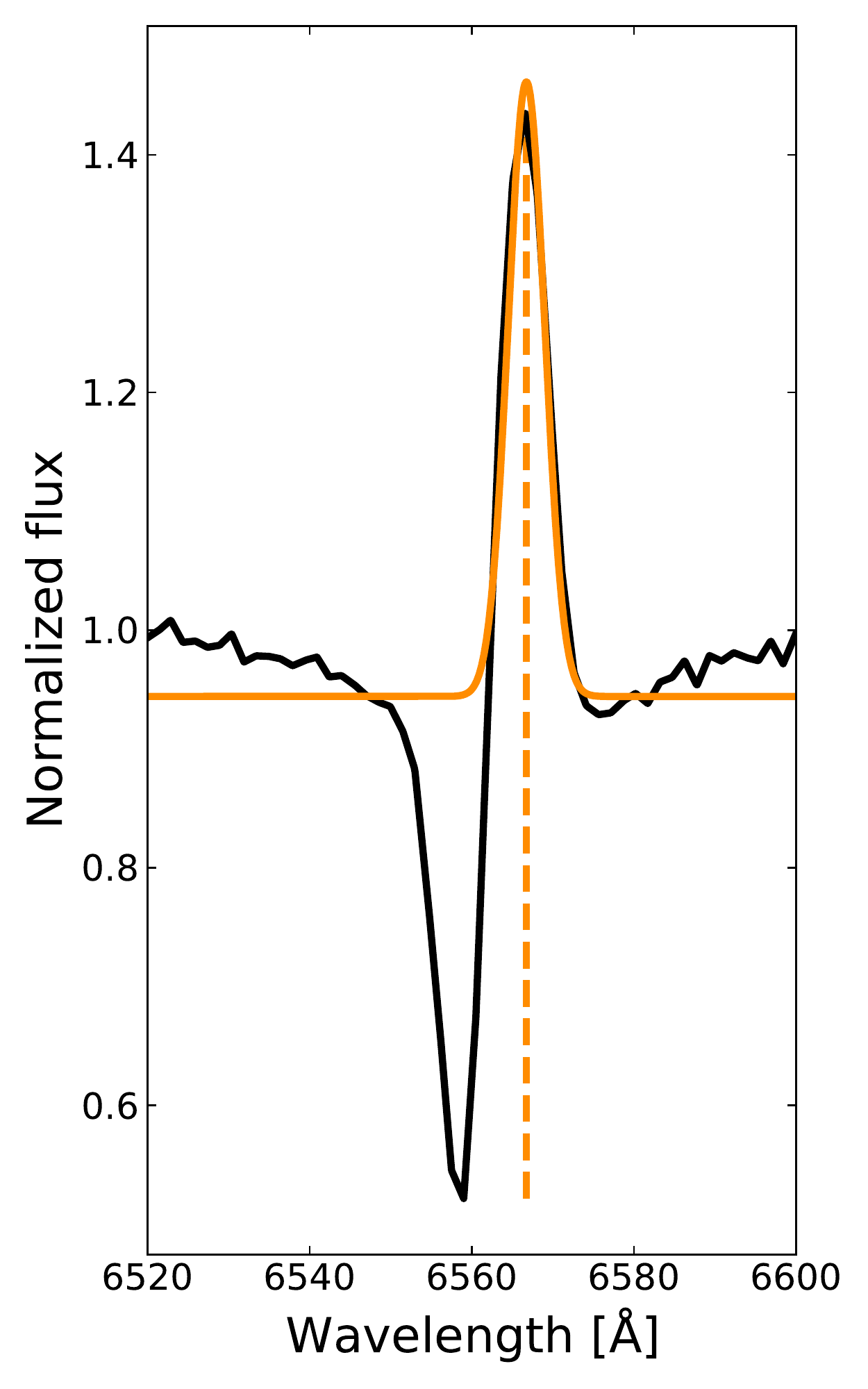}{0.2\textwidth}{Central wavelength determin}
          \fig{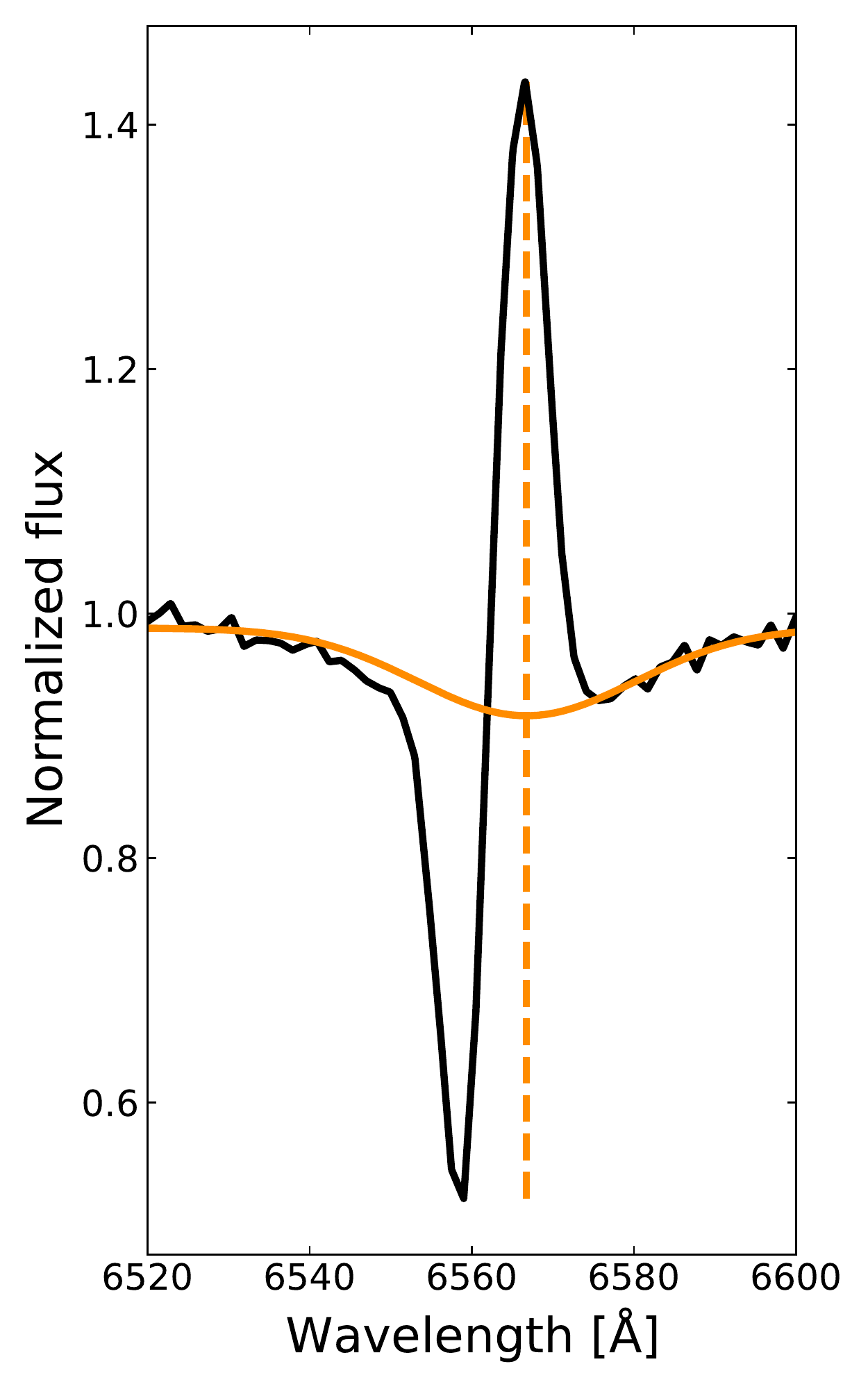}{0.2\textwidth}{Wide absorption fitting}
          \fig{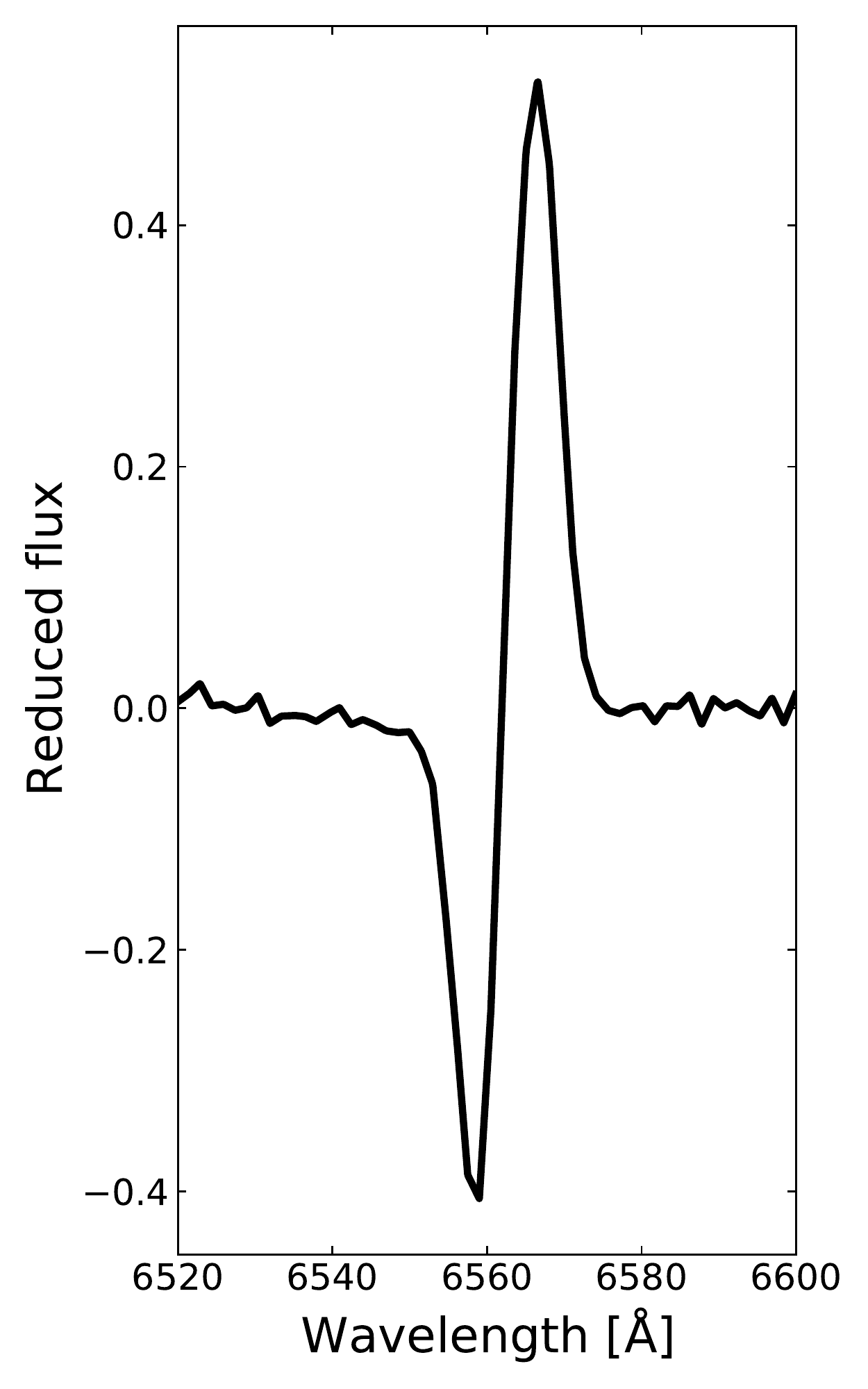}{0.2\textwidth}{Reduced flux}
          \fig{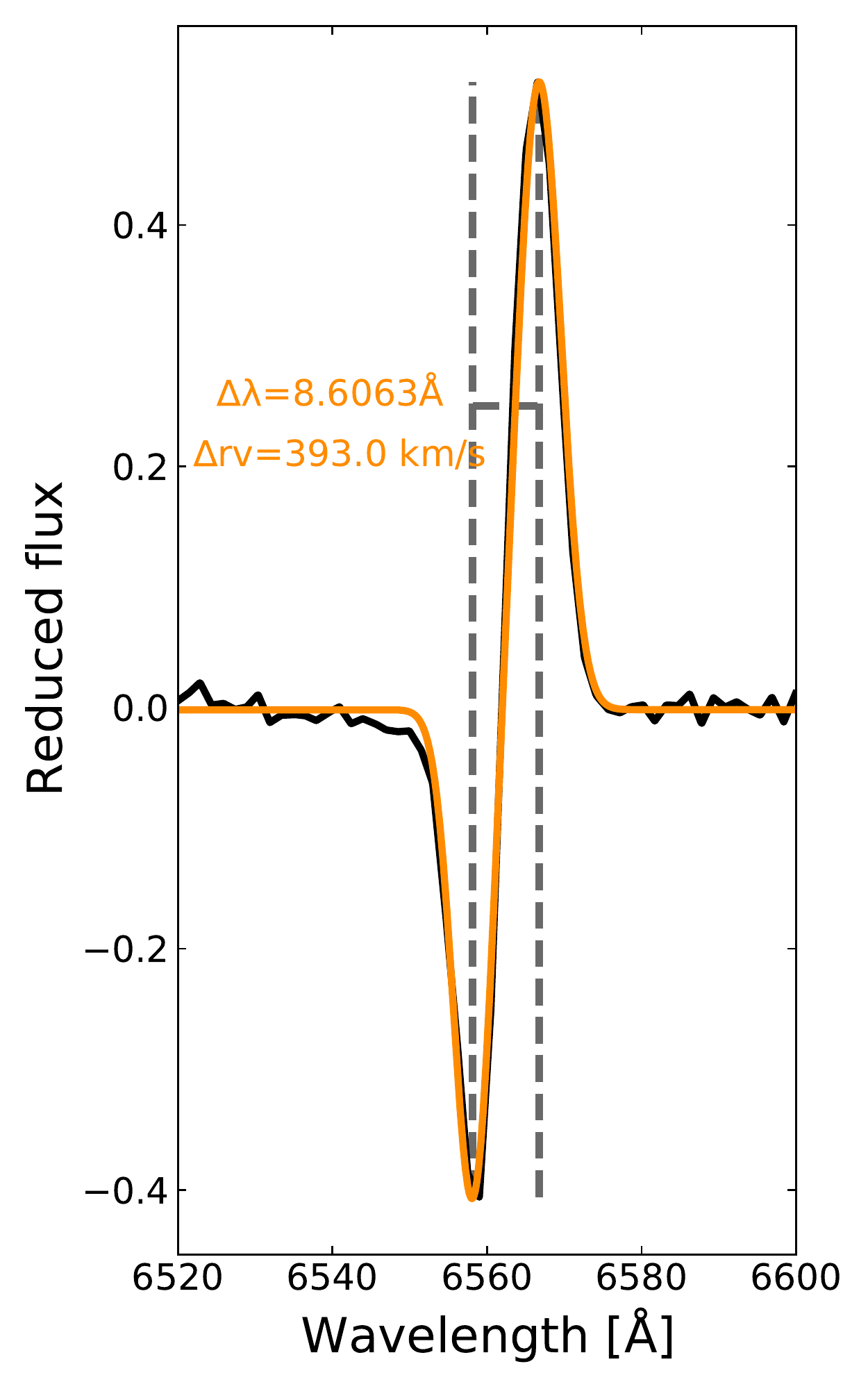}{0.2\textwidth}{Double Gaussian fitting}
          }

\caption{LAMOST J034753.05+291200.0. An example of stellar wind velocity estimation based on H$\alpha$ P Cygni profile with absorption profile.
\label{fig:fig3}}
\end{figure*}

Velocities of stellar winds have long been studied, such as \citet{AtlasPCyg.1979ApJS...39..481C,SEIPCygni1987ApJ...314..726L,PN.1989ApJ...345..339C,PCygni.1997A&A...326.1117N}. To calculate stellar wind or accretion flow velocities, the first process is to determine whether H$\alpha$ P Cygni profiles are accompanied by absorption profiles from central stars. We applied two different procedures to handle the spectra with or without absorption profiles. When it comes to a P Cygni or inverse P Cygni profile with photospheric absorption components, we need to remove the absorption components so as to get the pure P Cygni or inverse P Cygni. Figure \ref{fig:fig3} is an example of our approach. We used local fitting to determine the central wavelength of the emission line and using the same central wavelength to fit half of the wide absorption wing using Gaussian fitting. Subtracting the previous fitting flux of wide absorption from normalized flux, we got the final reduced flux. Fitting the reduced flux with double-gaussian function, we got central wavelengths of absorption and emission components respectively. Then, the velocities of stellar winds or accretion flows can be estimated with the Doppler equation shown as follows.


\begin{equation}
\triangle rv = \triangle \lambda / \lambda \times C
\end{equation}
$\triangle rv$ is the difference of radial velocities between surrounding envelope and central star. $\triangle \lambda$ is the difference of central wavelengths. $\lambda$ is the corresponding wavelength(H$\alpha$ or H$\beta$). $C$ represents the vacuum speed of light.

\begin{figure*}[ht!]

\gridline{\fig{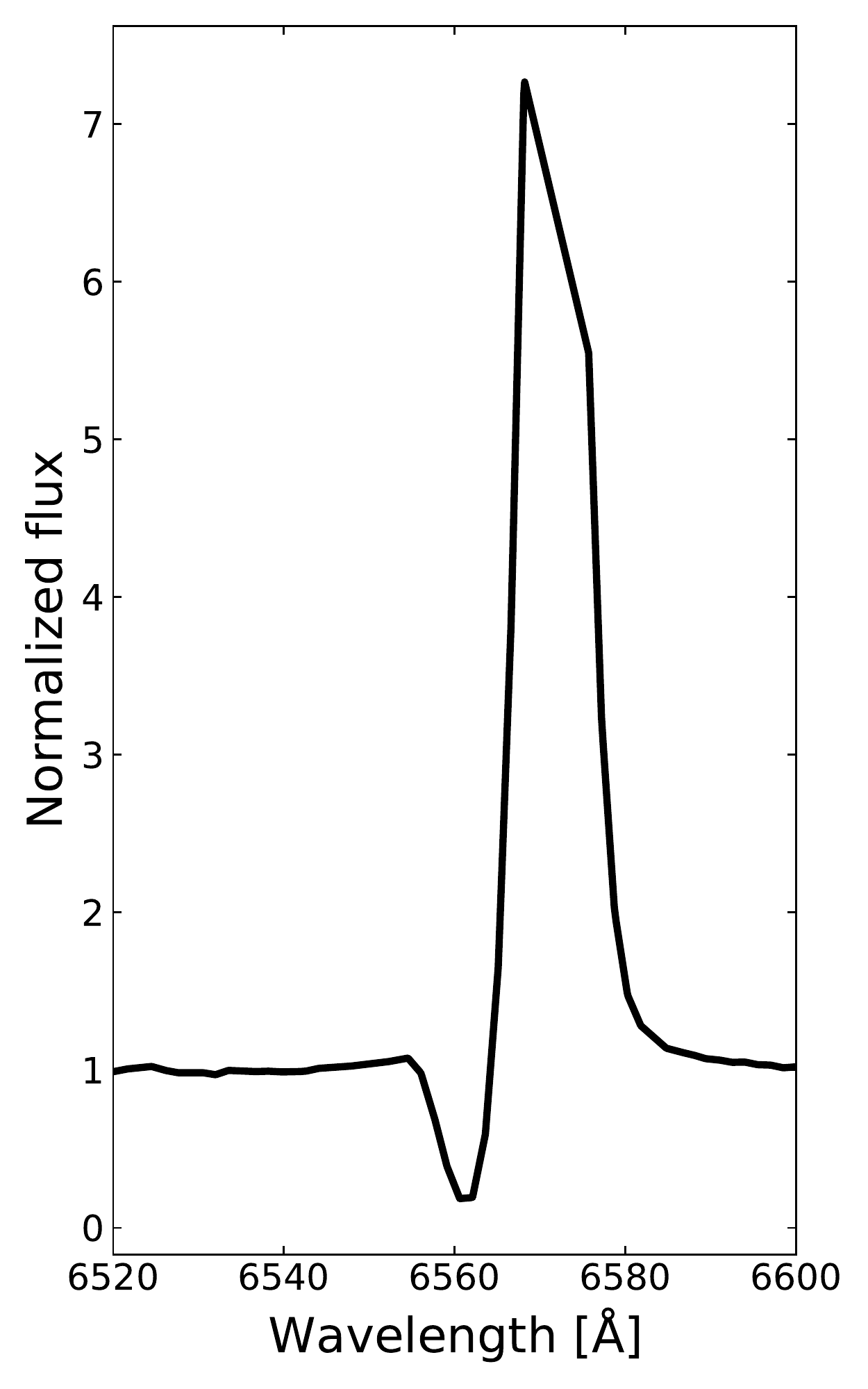}{0.2\textwidth}{Original H$\alpha$}
          \fig{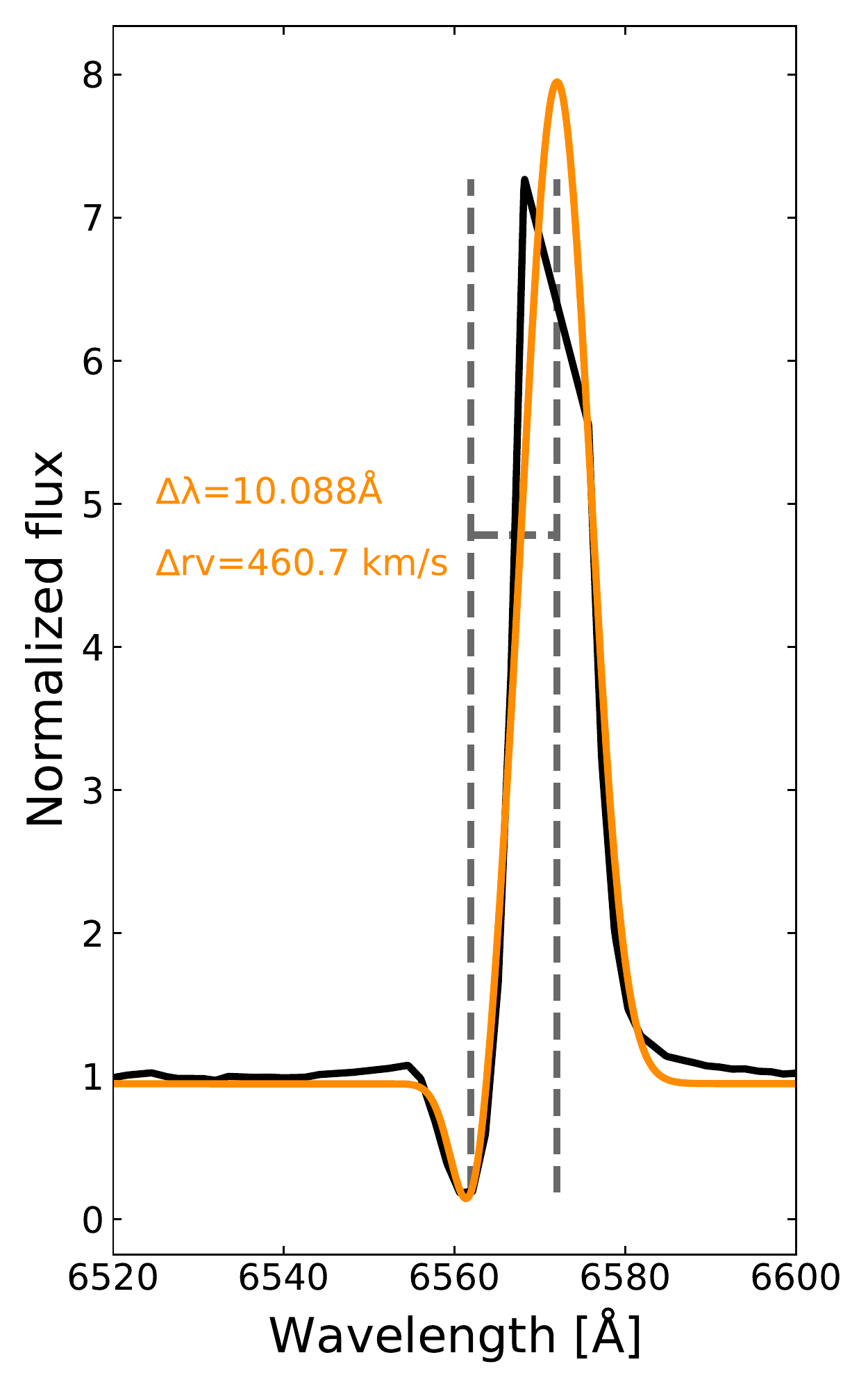}{0.2\textwidth}{Double Gaussian fitting of H$\alpha$}
          \fig{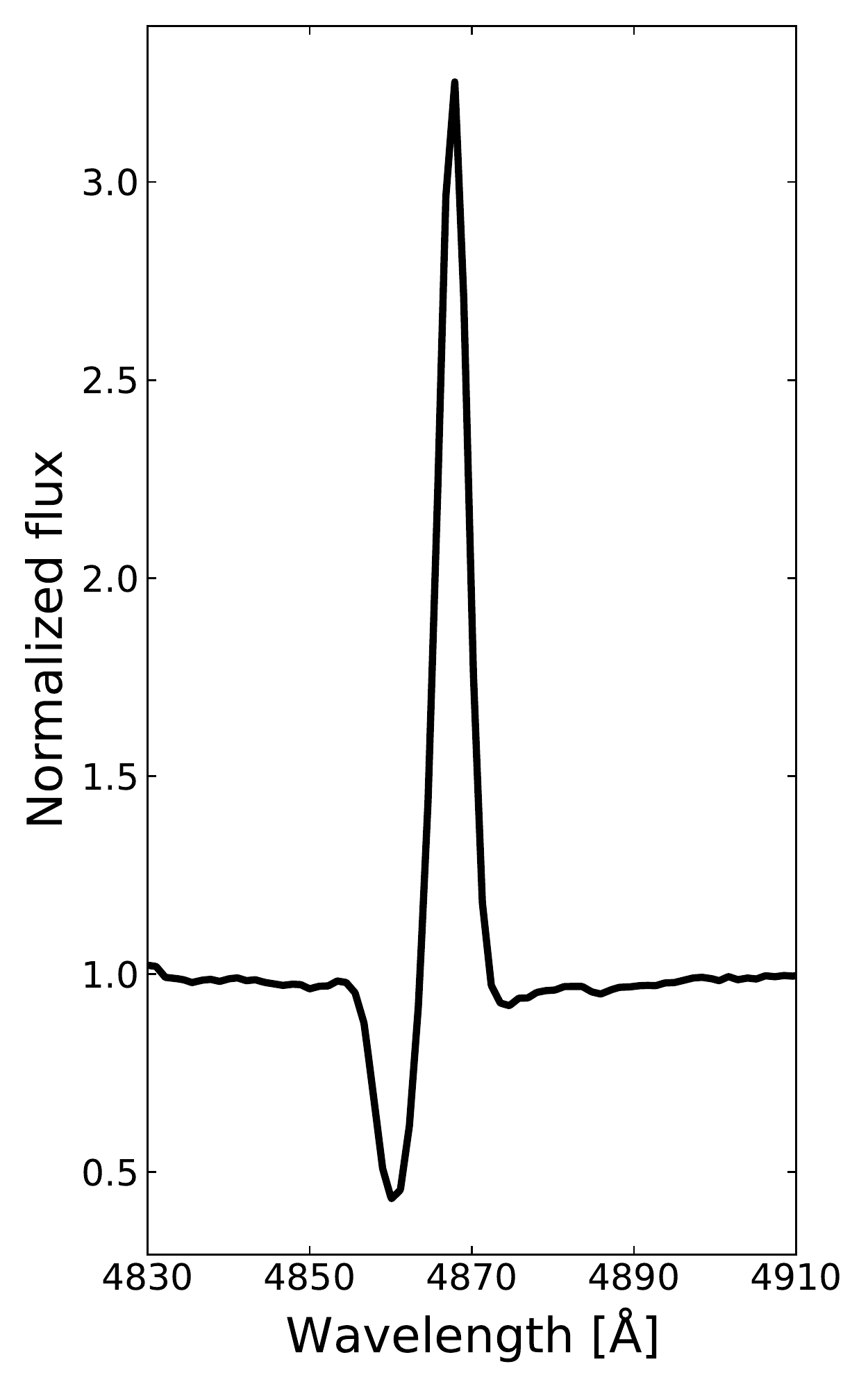}{0.2\textwidth}{Original H$\beta$}
          \fig{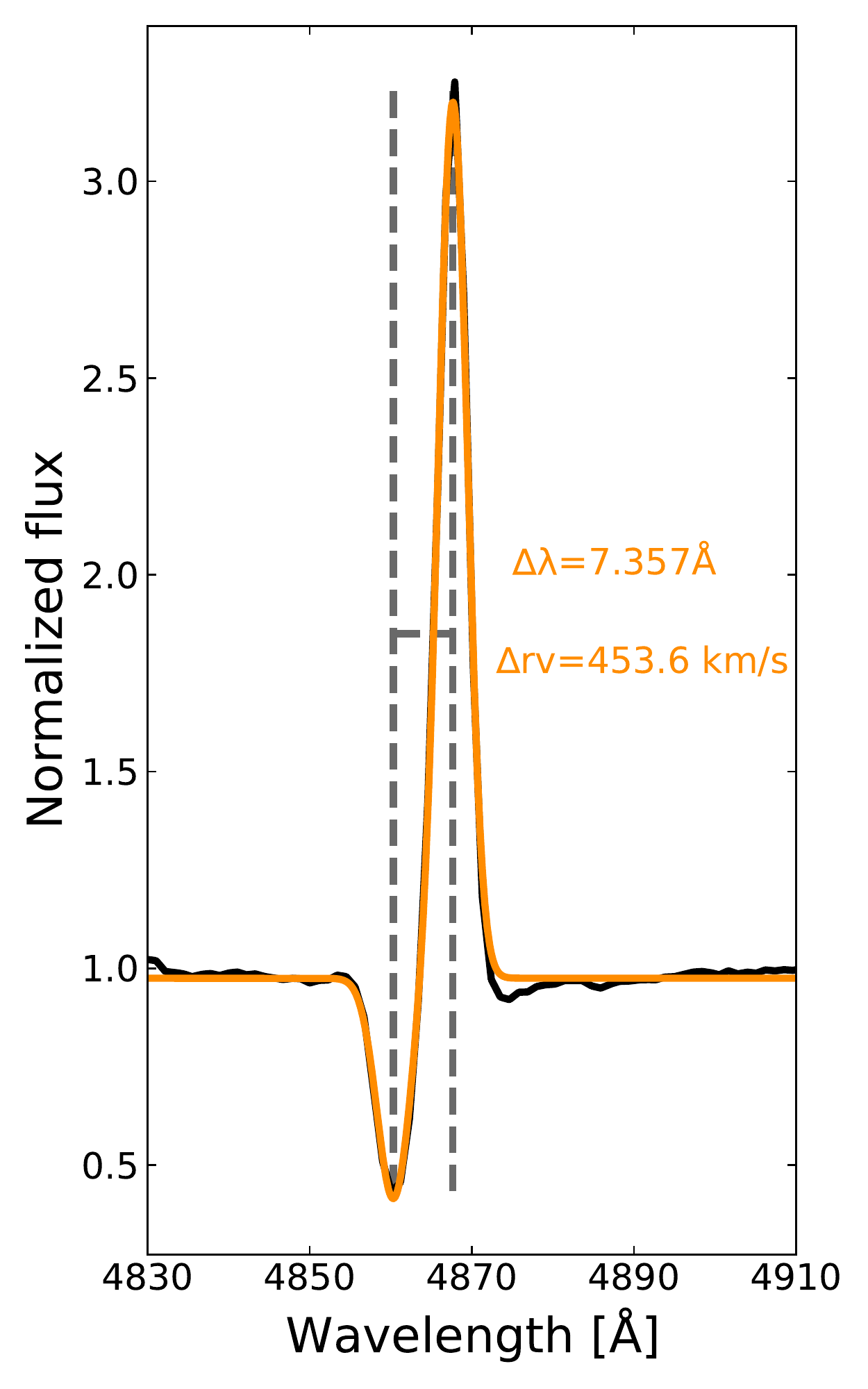}{0.2\textwidth}{Double Gaussian fitting of H$\beta$}
          }

\caption{LAMOST J035859.44+561112.6. An example of stellar wind velocity estimation based on P Cygni profiles without absorption profile. The H$\alpha$ and H$\beta$ emission profiles are both legible enough for curve fitting. The final estimation of stellar wind velocity is the mean of two speeds.
\label{fig:fig4}}
\end{figure*}

When it comes to a perfect H$\alpha$ P Cygni or inverse P Cygni profile without photospheric absorption component, we need to fit the profile with double-gaussian function and estimate the stellar wind or accretion flow speed directly. If the profile of H$\beta$ is also clear enough, we can also estimate the speed with H$\beta$ profile. Then the final speed is the mean of them. An example is shown in Figure \ref{fig:fig4}. According to the velocities estimated as shown in Table \ref{tab:velocities}, 74\% of targets in our samples have the speeds of stellar winds or accretion flows within $100\sim400$ km/s.

\section{Herbig and Classical Ae/Be candidates} \label{sec:HAeBe}

\subsection{Cross match with available catalogs} \label{subsec:crossmatch}
Before making a further analysis such as searching for new Herbig and Classical Ae/Be candidates, we cross-matched our sample with available catalogs of 4 specific types of stars, including Algol-type binary, cataclysmic variable star, CAeBe and HAeBe.

\subsubsection{Algol-type binary\label{subsubsec:algol}}

Algol variables or Algol-type binaries belong to a class of semidetached eclipsing binary systems. The primary component is an early type main sequence star without filling Roche lobe.The less massive secondary component with filling Roche lobe lies above the main-sequence in HR diagram. \citet{Algol1.2004A&A...417..263B} provided a catalog of 411 Algol-type binary stars and 1,872 Algol-type binary candidates. \citet{Algol2.2018ApJS..238....4P} presented an updated catalog of 4,680 northern eclipsing binaries with Algol-type light-curve morphology. We cross-matched our sample with the above catalogs of Algol-type binaries and got 33 spectra of 30 common sources.

\subsubsection{Cataclysmic variable star\label{subsubsec:CV}}

A Cataclysmic variable star (CV) is a close binary with a white dwarf primary and a mass transferring normal star secondary. The accretion disk around the primary is one of the formation mechanisms for H$\alpha$ emission line. \citet{CV.2020AJ....159...43H} presented a sample of 380 spectra of 245 cataclysmic variables based on the data of LAMOST DR5. We cross matched our sample with the CV catalog and got 20 spectra of 16 common sources.

\subsubsection{CAeBe and HAeBe\label{subsubsec:CAeBe and HAeBe}}

\citet{252HAeBe.2018A&A...620A.128V} proposed a catalog of 252 HAeBes gathered from previous works(\citet{HAeBe1.2016NewA...44....1C,HAeBe5.2015MNRAS.453..976F,HAeBe6.2013MNRAS.429.1001A,HAeBe7.2010A&A...517A..67C,HAeBe8.2010AJ....139...27S,HAeBe9.2006MNRAS.367..737B,HAeBe10.2006ApJ...653..657M,HAeBe2.2006Ap&SS.305...11Z}, \citet{JHK.2005AJ....129..856H,HAeBe4.2003AJ....126.2971V,HAeBe3.1994A&AS..104..315T}). \citet{BeSS.2011AJ....142..149N} provided 2256 CBes and 55 HAeBes. We also gathered 159 CAes provided by \citet{2021MNRAS.501.5927A}. Then we cross-matched our sample with the above known HAeBes and CAeBes and got 269 spectra of 181 known CAeBes along with 12 spectra of 9 known HAeBes.

\subsubsection{Cross matching with \citeauthor{2016RAA....16..138H}\label{subsubsec:DR2}}

We also cross-matched our sample with the catalog of \citet{2016RAA....16..138H} including 11,204 emission-line spectra from LAMOST DR2. As we know the later data releases of LAMOST should include all previous releases, which means all observed spectra will be accessed in the latest release. However, the data reduction or classification pipeline is continuously updated. As a result, the reduced spectra or classes for a small fraction of objects might be different. The cross-matched result shows that 8,149 out of 11,204 spectra are included in our samples while 3,055 spectra are not. We checked the 3,055 spectra in DR7 of LAMOST and found 1,073 of them are classified as F or later type stars which were classified as A-type in DR2. For the remaining 1,982 entries in \citeauthor{2016RAA....16..138H} that not be included in our sample,  we visually inspected the spectra of LAMOST DR7 and found none of them shows any evident emission-line feature. The reason is that the sky subtraction algorithm in the data reduction pipeline has been upgraded since LAMOST DR3 according to \citet{sky.2017RAA....17...91B}. Thus we confirmed that these 3,055 spectra in DR7 will not be compiled into our catalog.


\subsection{Collected CAeBe and HAeBe in literature \label{subsec:AeBe in literaturer}}
Apart from stars with spectra contaminated by nebula emission  lines and close binaries such as CVs and Algol-type binaries, the remainders are mainly CAeBes and HAeBes.
Based on the fact that IR excesses of HAeBes are much stronger than that of CAeBes due to re-radiation caused by dust envelopes covering central stars, people used to search for HAeBe with IR photometric data. \citet{1984A&AS...55..109F} suggested an empirical method using (H-K,K-L) diagram. Considering the lack of L band, \citet{2016RAA....16..138H} replaced L band with W1, the first band of WISE. \citet{JHK.2005AJ....129..856H} provided another empirical method using (H-K,J-H) diagram with the HAeBes of \citet{2004AJ....127.1682H}. \citet{2021MNRAS.501.5927A} presented a catalog of 159 new CAes using empirical (H-K,J-H) diagram based on LAMOST DR5(\citet{2019yCat.5164....0L}).

Lacking enough known samples when the methods above proposed, the separation between HAeBes and CAeBes are quite rough and incomplete based on either of the above empirical IR color criteria. Here in this paper, we proposed an updated criterion for separating HAeBes from CAeBes based on the up-to-date known HAeBes and CAeBes. We gathered 252 HAeBes from the catalog of \citet{252HAeBe.2018A&A...620A.128V}, 2,256 CBes and 55 HAeBes (53 of which are included in previous catalog) from the catalog of \citet{BeSS.2011AJ....142..149N}, and 159 CAes presented by \citet{2021MNRAS.501.5927A}. The final known data set for making an update criterion is shown as Table \ref{tab:composition}.

\begin{deluxetable}{cccc}
\tablenum{4}
\tablecaption{Collected CAeBe and HAeBe in literature\label{tab:composition}}
\tablewidth{0pt}
\tablehead{
\colhead{Classical Be} & \colhead{Herbig Ae/Be} & \colhead{Classical Ae}
}
\startdata
2244 & 254 & 159
\enddata
\end{deluxetable}

\subsection{Interstellar extinction correction\label{subsec:interstellar}}

In order to use color-color diagrams to distinguish HAeBes from CAeBes, we retrieved magnitudes of J, H, and K from 2MASS \citep{2003yCat.2246....0C} and that of W1 from WISE \citep{2010AJ....140.1868W}. There are three main methods to correct interstellar extinction, intrinsic color method \citep{1970A&A.....4..234F}, 2D dust map \citep{1998ApJ...500..525S}, and 3D dust maps\footnote{\url{http://argonaut.skymaps.info/}}. The Intrinsic color method needs precise spectral types. The interstellar extinction given by 2D dust map is usually larger than the real value. 3D dust maps have three versions based on the extinction relations of \citet{Extinction.1999PASP..111...63F,Extinction.1989ApJ...345..245C}, that is Bayestar19, Bayestar17, and Bayestar15 \citep{Bayestar19.2019ApJ...887...93G,Bayestar17.2018MNRAS.478..651G,Bayestar15.2015ApJ...810...25G}.
Among these versions, the unit of Bayestar15 reported using the same units as SFD \citep{1998ApJ...500..525S}. So we adopted coefficients using the same unit as SFD in literature, for example, Table 6 of \citet{SFD.2011ApJ...737..103S} and Table 1, 2 of \citet{W1.2013MNRAS.430.2188Y}. We referenced  the later and adopted $R_{(J-H)}=0.26 \pm 0.005$, $R_{(H-K)}=0.16 \pm 0.006$ and $R_{(K-W1)}=0.12 \pm 0.008$.


\subsection{Discriminate criteria \label{sec:criterion}}

\begin{figure*}[ht!]

\gridline{\fig{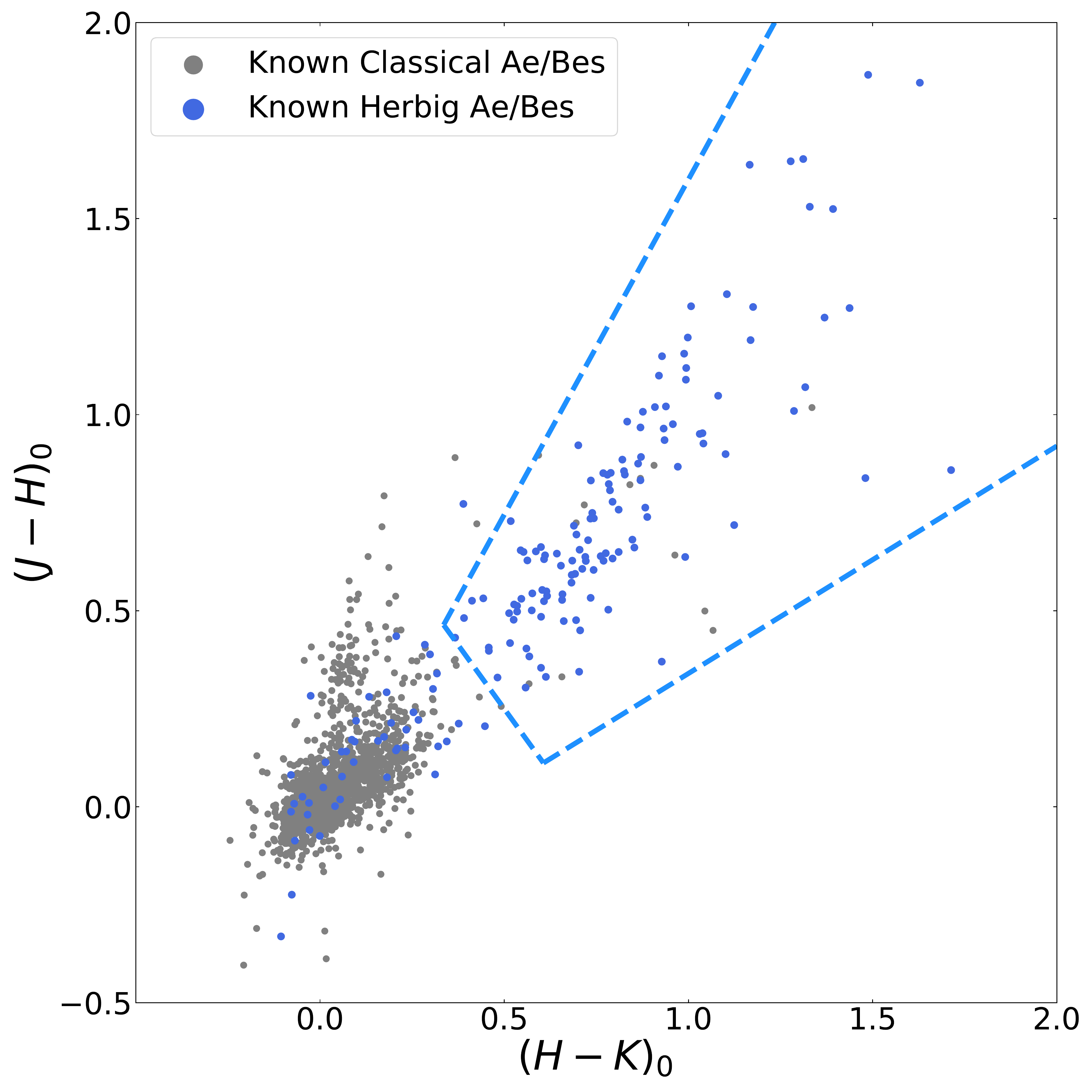}{0.5\textwidth}{Known HAeBes and CAeBes on (H-K,J-H) diagram}
          \fig{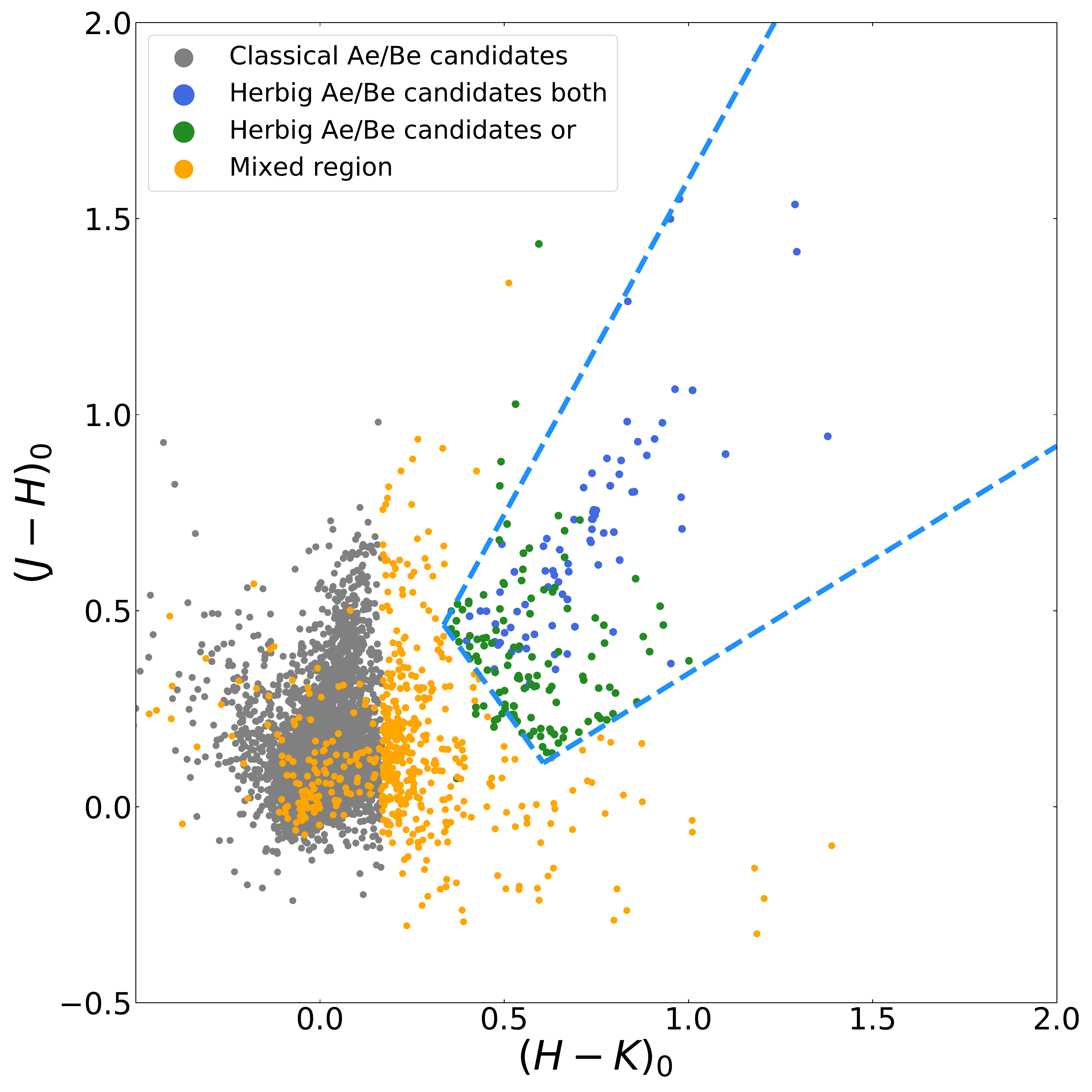}{0.5\textwidth}{HAeBe and CBe candidates in LAMOST DR7}
          }
\gridline{\fig{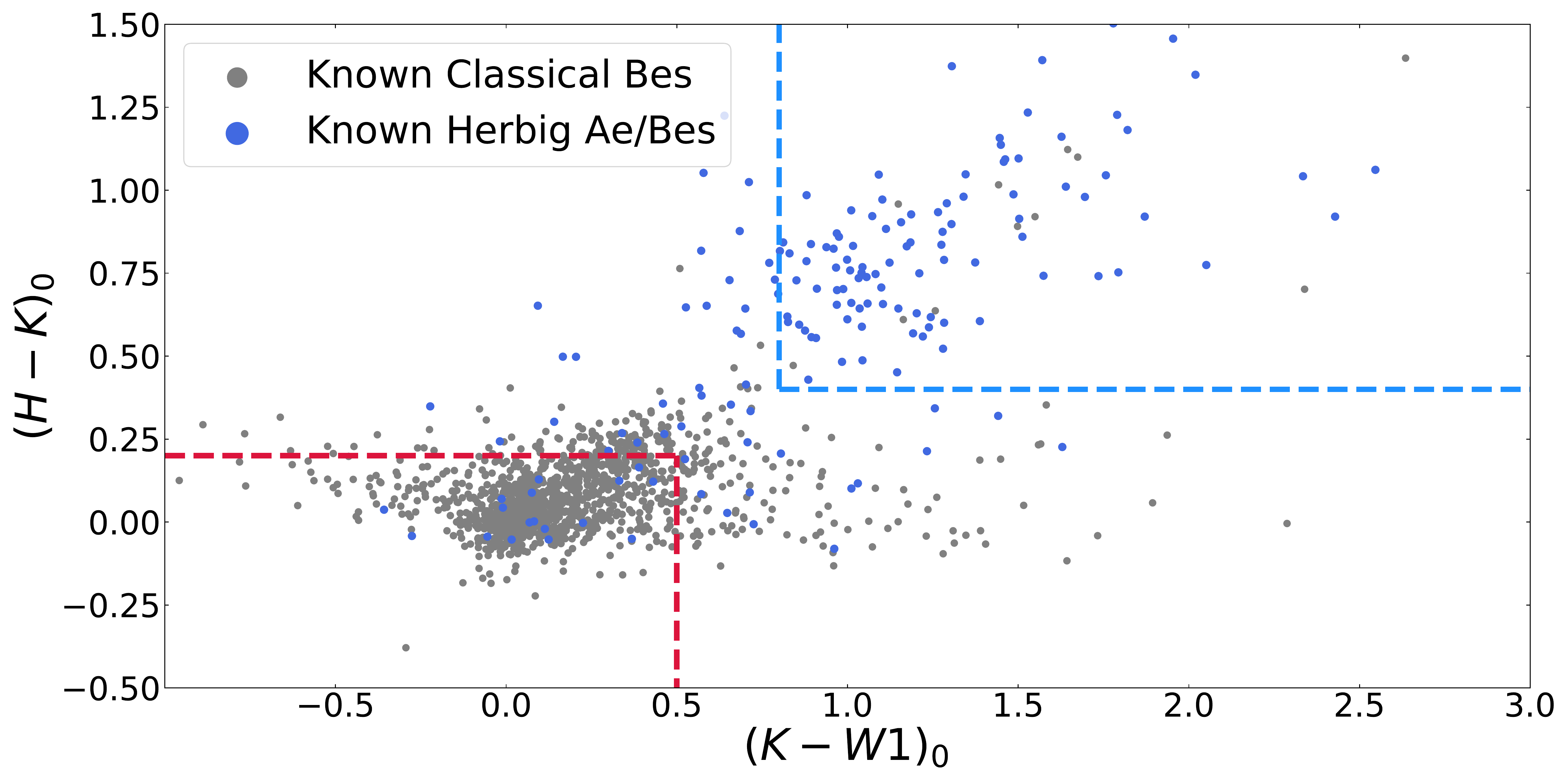}{0.5\textwidth}{Known HAeBes and CAeBes on (H-K,K-W1) diagram}
          \fig{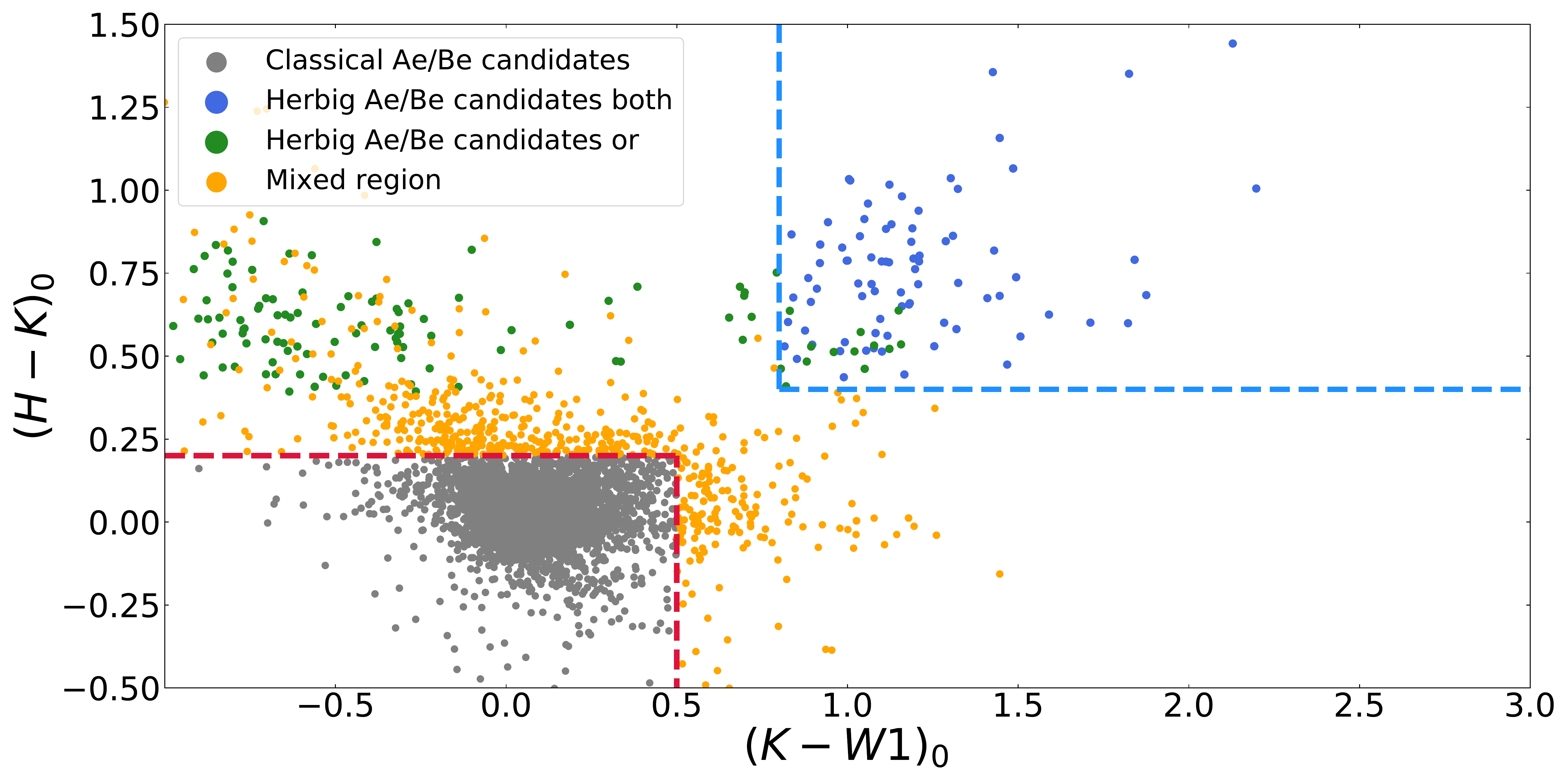}{0.5\textwidth}{HAeBe and CAeBe candidates in LAMOST DR7}
          }
\caption{The upper two panels are ($\rm(H-K)_0$,$\rm(J-H)_0$) diagrams (colors have been transformed into CIT system). Two bottom panels are ($\rm(K-W1)_0$,$\rm(H-K)_0$) diagrams. The interstellar extinction of all the colors have been corrected based on 3d dust map. The two left panels are the distribution of known Herbig and classical Ae/Bes on two color-color diagrams. Gray solid circles represent known CAeBes and blue solid circles are known HAeBes. The dividing lines represent two individual criteria. The two right sub-figures show the result after applying the new proposed criterion to our sample. Gray solid circles represent CAeBe candidates. Blue and green solid circles are HAeBe candidates. Blue ones are the common candidates through two independent criteria while green ones are only recognized as HAeBe candidates in either one diagram. Yellow solid circles locate in the mixed region where maybe either classical or Herbig Ae/Bes.
\label{fig:fig5}}
\end{figure*}

When plotted on the (H-K,J-H) and (H-K,K-W1) diagrams as shown in the left two panels of Figure \ref{fig:fig5}, known HAeBes and CAeBes are mainly centralized in different regions. When plotted on (H-K,J-H) diagram, color magnitudes have been transformed into CIT system with the following equations proposed by \citet{CIT.2001AJ....121.2851C}.

\begin{equation}
(J-H)_{2MASS} = (1.076 \pm 0.010)(J-H)_{CIT} + (-0.043 \pm 0.006)
\end{equation}

\begin{equation}
(H-K)_{2MASS} = (1.026 \pm 0.020)(H-K)_{CIT} + (0.028 \pm 0.005)
\end{equation}

In the left bottom panel of Figure \ref{fig:fig5}, we adopt the criterion proposed by \citet{2016RAA....16..138H} that HAeBes typically locate in the region of $H-K > 0.4$, $K-W1 > 0.8$ while CAeBes typically locate in the region of $H-K < 0.2$, $K-W1 < 0.5$. In the left upper panel of Figure \ref{fig:fig5}, we proposed a set of inequations to determine the location of HAeBes shown bellow. We took the combination of HAeBes got from two diagrams separately as final HAeBe candidates while the CAeBes determined by (H-K,K-W1) diagram as the final CAeBe candidates.


\begin{equation}
(J-H)_0 > 1.71 \times (H-K)_0 - 0.11
\end{equation}

\begin{equation}
(J-H)_0 > 0.58 \times (H-K)_0 - 0.24
\end{equation}

\begin{equation}
(J-H)_0 > -1.3 \times (H-K)_0 + 0.9
\end{equation}

\subsection{Feasibility analysis of the proposed criterion \label{sec:feasibility analysis}}

Before applying these new criteria to our sample, feasibility analysis is essential. We applied our new criterion on known Herbig and classical Ae/Bes, and the prediction result is shown in Table \ref{tab:prediction}. Considering the lack of necessary photometric magnitudes of several sources, the final quantity of samples can be predicted is 1,475, including 171 HAeBes and 1,299 CAeBes.

\begin{deluxetable*}{ccccccc}
\tablenum{5}
\tablecaption{Prediction result of known HAeBes and CAeBes\label{tab:prediction}}
\tablewidth{0pt}
\tablehead{
\colhead{} & \colhead{} & \colhead{Classified as} & \colhead{Classified as} & \colhead{Classified in} & \colhead{} & \colhead{}\\
\colhead{} & \colhead{Sample quantity} & \colhead{Herbig Ae/Be} & \colhead{classical Ae/Be} & \colhead{mixed region} & \colhead{precision} & \colhead{recall}
}
\startdata
Herbig Ae/Be & 171 & 127 & 17 & 27 & 90.714\% & 74.269\%\\
classical Ae/Be & 1299 & 13 & 1012 & 274 & 98.348\% & 77.906\%\\
\enddata
\tablecomments{This table shows the prediction result of newly proposed criterion on known Herbig and classical Ae/Bes. }
\end{deluxetable*}

\subsection{Application to emission-line stars \label{sec:application}}

As is well-known that H\Rmnum{2} regions and other regions with molecular clouds diffused are the possible harbors of star formation. We should include these stars with spectra contaminated by nebula emission  lines when searching for young stellar objects(YSOs) such as HAeBes and TTs. However, when it comes to search for CAeBes, it's not wise to take into account of spectra contaminated by nebula emission  lines for consideration of candidate purity. We selected the emission-line stars with explicit emission-line profiles, without contaminated by nebula emission  lines as possible CAeBe data set. We applied the above criterion to our sample and got 235 spectra of 201 HAeBe candidates along with 6,741 spectra of 5,547 CAeBe candidates. The right two panels in Figure \ref{fig:fig5} show the distribution of candidates on two color-color diagrams.






\begin{figure*}
\begin{center}
\includegraphics[width=\textwidth]{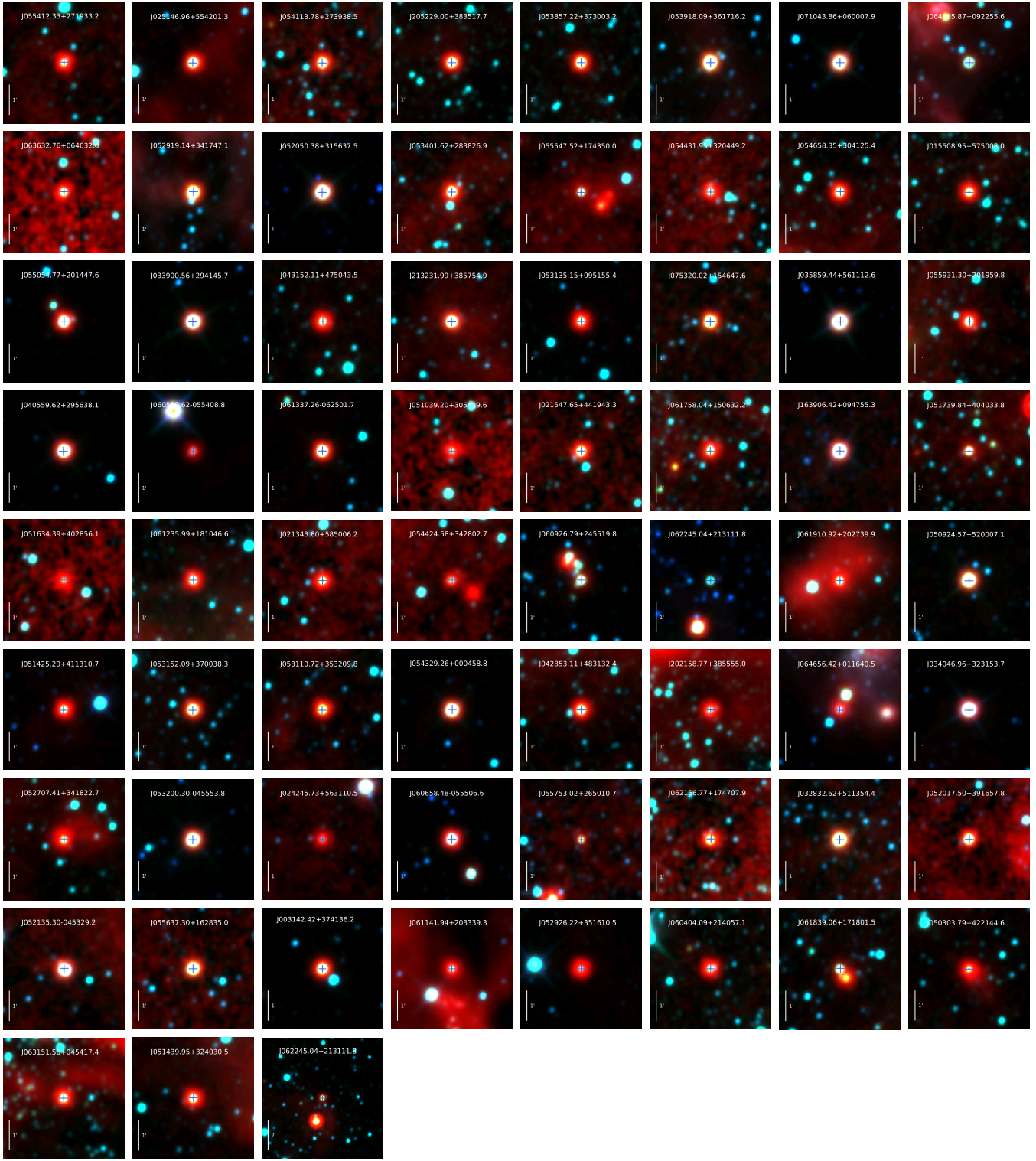}
\end{center}
\caption{WISE pseudo-color images atlases of 66 confirmed HAeBes (W1 in blue, W2 in green, and W4 in red).  The target is marked with a blue cross in the middle of each image. The former 66 images are cut as 4'$\times$4'. The last one is the same source of ’J062245.04+213111.8’ (the same target of line 5 column 6) zoomed out to a larger field  8'$\times$8' because the W4 excess of the target can not be seen in the 4'$\times$4' image caused by the color depth scale.
\label{fig:fig6}}
\end{figure*}

\begin{figure*}
\begin{center}
\rotatebox{90}{\includegraphics[width=9.2in]{spectra.pdf}}
\end{center}
\caption{H$\alpha$ profiles of the 66 confirmed Herbig Ae/Bes. It should be note that the spectra of source J034046.96+323153.7 has been masked on in the H$\alpha$ region, while the target have medium resolution spectrum showing H$\alpha$ emission, and the medium resolution spectrum can be seen in Figure \ref{fig:fig8} .
\label{fig:fig7}}
\end{figure*}

\begin{deluxetable*}{ccccccc}
\tablenum{6}
\tablecaption{66 confirmed HAeBes\label{tab:confirmed}}
\tablewidth{0pt}
\tablehead{
\colhead{ } & \colhead{ } &  \colhead{ra} & \colhead{dec} & \colhead{Distance} & \colhead{ } & \colhead{ }\\
\colhead{Designation} & \colhead{Simbad\_ID} & \colhead{[deg]} & \colhead{[deg]} & \colhead{[pc]} & \colhead{Medium-resolution} & \colhead{LAMOST\_class}
}
\startdata
J003142.42+374136.2 & IRAS 00290+3725 & 7.9267569 & 37.6934014 & 1086.05-1309.89 &   & A6IV\\
J015508.95+575008.0 &    & 28.78732 & 57.835581 & 1583.05-1813.44 & Y & A5V\\
J021343.60+585006.2 &    & 33.431667 & 58.835083 & 1773.75-2293.8 &   & A5V\\
J021547.65+441943.3 &    & 33.948563 & 44.328719 & 3307.85-4671.33 &  & A1V\\
J024245.73+563110.5 &    & 40.690549 & 56.5196 & 2074.72-2562.3 &    & A2V\\
J025146.96+554201.3 & 2MASS J02514696 & 42.945683 & 55.70038 & 2480.88-2756.38 &   & A2V\\
                    & +5542014 &   &   &   &   &   \\
J032832.62+511354.4 &    & 52.135946 & 51.231794 & 1829.78-2119.7 &    & A1IV\\
J033900.56+294145.7* & V* V1185 Tau & 54.752343 & 29.696032 & 387.19-402.89 &    & A2IV\\
J034046.96+323153.7* & HD 278937 & 55.19567 & 32.53159 & 298.1-312.71 & Y & A6IV
\enddata
\tablecomments{* means source is a known HAeBe. The 'LAMOST\_class' column is the result of LAMOST 1D-pipeline. The 'Distance' column is provided by \citet{Distance.2018AJ....156...58B}. Complete list is available in electronic form.}
\end{deluxetable*}

\begin{deluxetable*}{ccccccccc}
\tablenum{7}
\tablecaption{HAeBes associated with solar-nearby SFRs\label{tab:nebulae}}
\tablewidth{0pt}
\tablehead{
\colhead{ } & \colhead{ } &  \colhead{l} & \colhead{b} & \colhead{Distance} & \colhead{ } & \colhead{Distance\_SFR} & \colhead{l\_SFR} & \colhead{b\_SFR}\\
\colhead{Designation} & \colhead{Simbad\_ID} & \colhead{[deg]} & \colhead{[deg]} & \colhead{[pc]} & \colhead{SFR} & \colhead{[pc]} & \colhead{[deg]} & \colhead{[deg]}
}
\startdata
J033900.56+294145.7* & V* V1185 Tau & 161.18303 & -20.46282 & 387.19-402.89 & Per OB2 & 318 & 160±4 & -17.5±4.5\\
J034046.96+323153.7* & HD 278937 & 159.59553 & -18.02591 & 298.1-312.71 & Per R1 & 400 & 158.5±2.5 & -19.5±2.5\\
J040559.62+295638.1 & IRAS 04028+2948 & 165.61035 & -16.3794 & 336.54-345.18 & LDN1498 & 140 & 169.97 & -19.26\\
J052135.30-045329.2 &    & 206.87517 & -22.19043 & 365.47-373.68 & Ori OB1 & 500 & 206±9 & -19±7\\
J053135.15+095155.4 & 2MASS J05313515 & 194.65357 & -12.78158 & 376.4-581.07 & LDN1582 & 460 & 192.21 & -11.56\\
                    &       +0951553  &           &           &              &         &     &        &       \\
J053200.30-045553.8* & V* UY Ori & 208.18504 & -19.90146 & 342.63-363.07 & Ori R2 & 500 & 210.5±2.5 & -19.5±1.5\\
J054329.26+000458.8 & V* GT Ori & 204.9259 & -15.03216 & 412.14-431.77 & Ori R1 & 500 & 205.5±1.5 & -15.5±2.5\\
J060559.62-055408.8 &    & 213.05964 & -12.78429 & 731.23-798.86 & Ori OB1 & 500 & 206±9 & -19±7\\
J060658.48-055506.6* & 2MASS J06065848 & 213.18603 & -12.57318 & 809.39-878.68 & Ori OB1 & 500 & 206±9 & -19±7\\
                    &        -0555066 &           &           &              &         &     &        &       \\
J061337.26-062501.7* & 2MASS J06133726 & 214.39525 & -11.31698 & 806.41-861.77 & LDN1646 & 830 & 213.2 & -12.73\\
                    &        -0625017 &           &           &              &         &     &        &       \\
J064105.87+092255.6* & HD 261941 & 203.40713 & 1.99034 & 719.25-794.41 & Mon R1 & 715 & 202.5±1.5 & 1±2
\enddata
\tablecomments{* means source is a known HAeBe. 'SFR' column is the name of star formation region. LDN is the abbreviation for the Lynds clouds(\citet{nebulae4.1962ApJS....7....1L}).
}
\end{deluxetable*}

\section{Confirmation of new HAeBes } \label{sec:confirmation}

\citet{Herbig.1960ApJS....4..337H} originally defined HAeBes by the following criteria.

\begin{enumerate}
\item The spectral type is A or earlier, with emission lines.
\item The star lies in an obscured region.
\item The star illuminates fairly bright nebulosity in its immediate vicinity.
\end{enumerate}

\citet{Herbig.1972ApJ...173..353S} made the further definition including properties of infrared excess, time variation, linear polarization, and association with star-forming regions (SFRs). \citet{1984A&AS...55..109F} made a specific spectral analysis of 57 HAeBes and candidates. \citet{HAeBe7.2010A&A...517A..67C} confirmed 13 new HAeBes via high-resolution optical spectroscopy.

In this work, we try to use both LAMOST spectra and WISE photometric data to identify HAeBe since they have wider H$\alpha$ emission lines and strong excesses at 12 and 22$\mu m$. The WISE mapped the whole sky in four infrared bands W1,W2,W3,W4 centered at 3.4,4.6,12, and 22$\mu m$ \citep{WISE.2010AJ....140.1868W}.  The 3.4 and 12$\mu m$ filters include prominent PAH emission features, the 4.6$\mu m$ filter measures the continuum emission from very small grains, and the 22$\mu m$ filter sees both stochastic emission from small grains and the Wien tail of thermal emission from large grains. We have got 235 spectra of 201 HAeBe candidates with LAMOST DR7 low-resolution spectra and photometric magnitudes of 2MASS and WISE previously. We analysed the circumstellar environment of each candidate and confirmed 66  of them as HAeBes, shown in Table \ref{tab:confirmed}, based on the existence of strong excesses of W3 and W4 caused by dust envelopes/disks tightly around the central stars.
The difference between dust disks and molecular clouds are explicit in images of W3 and W4 band.  In the image of W3 and W4, the molecular clouds are large spreading background and the dust envelopes/disks show additional explicit IR excesses surrounding the central stars.
We used W1(blue), W2(green), W4(red) of WISE image atlases to generate RGB images\footnote{\url{https://irsa.ipac.caltech.edu/applications/wise/}} as shown in Figure \ref{fig:fig6}. The H$\alpha$ profiles of the 66 confirmed HAeBes are shown in Figure \ref{fig:fig7}.

\subsection{Associated with solar-nearby SFRs \label{sec:SFR}}

To establish the association of HAeBes with nearby SFRs, we employed the solar-nearby SFRs with distances provided by \citet{HAeBe7.2010A&A...517A..67C} (gathered from \citet{nebulae1.1966AJ.....71..990V,nebulae2.1999AJ....117..354D,nebulae3.1995A&AS..113..325H}). We finally got 11 HAeBes located in the known solar-nearby SFRs as shown in Table \ref{tab:nebulae}, 5 of which are newly identified. Besides, almost all of 66 HAeBes are located in the projection of molecular clouds despite the distance of clouds have not been measured.

\subsection{HAeBes with medium-resolution spectra \label{sec:medium-resolution}}
\begin{figure}
\gridline{\fig{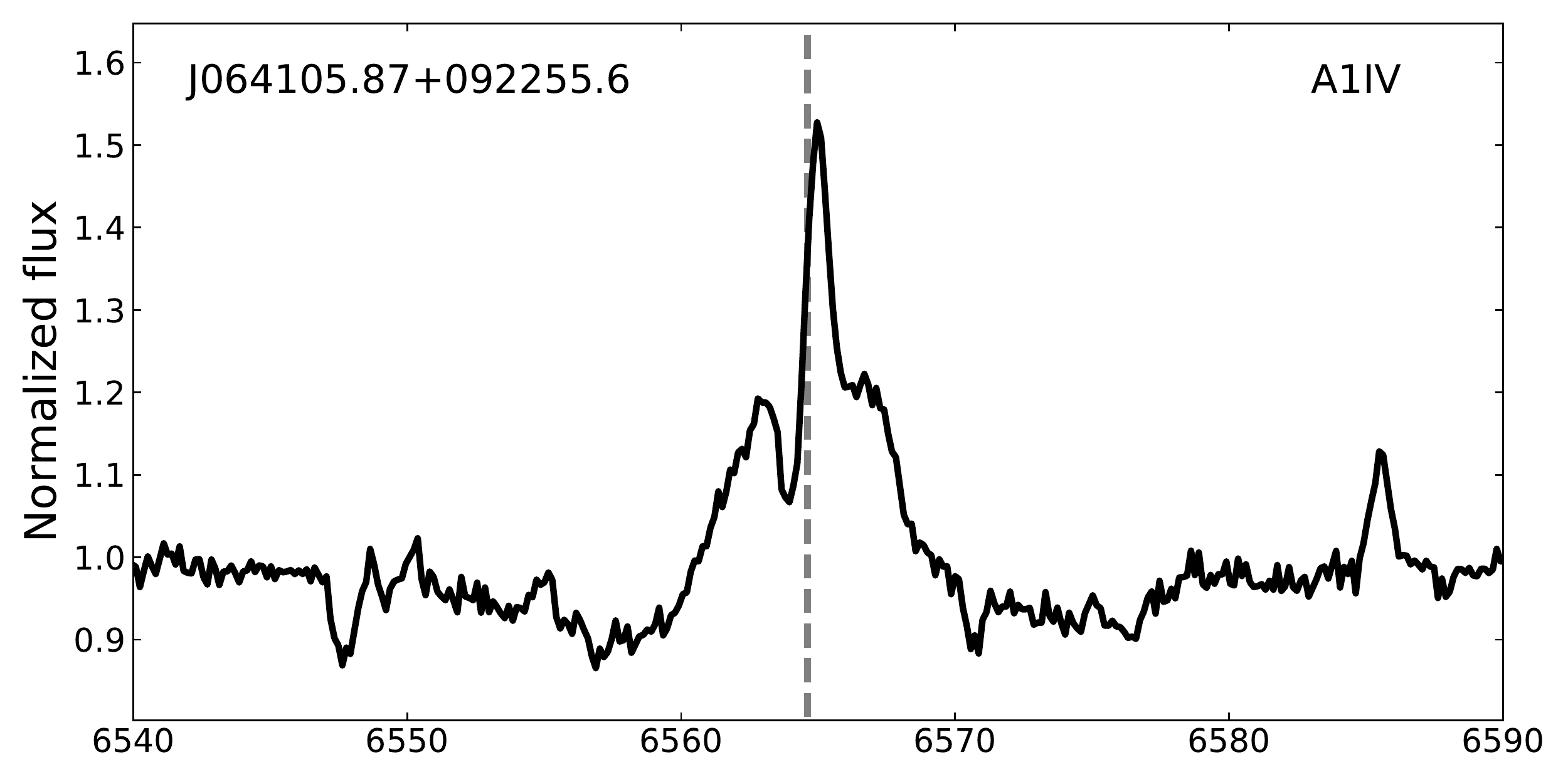}{0.5\textwidth}{J064105.87+092255.6}
          }
\gridline{\fig{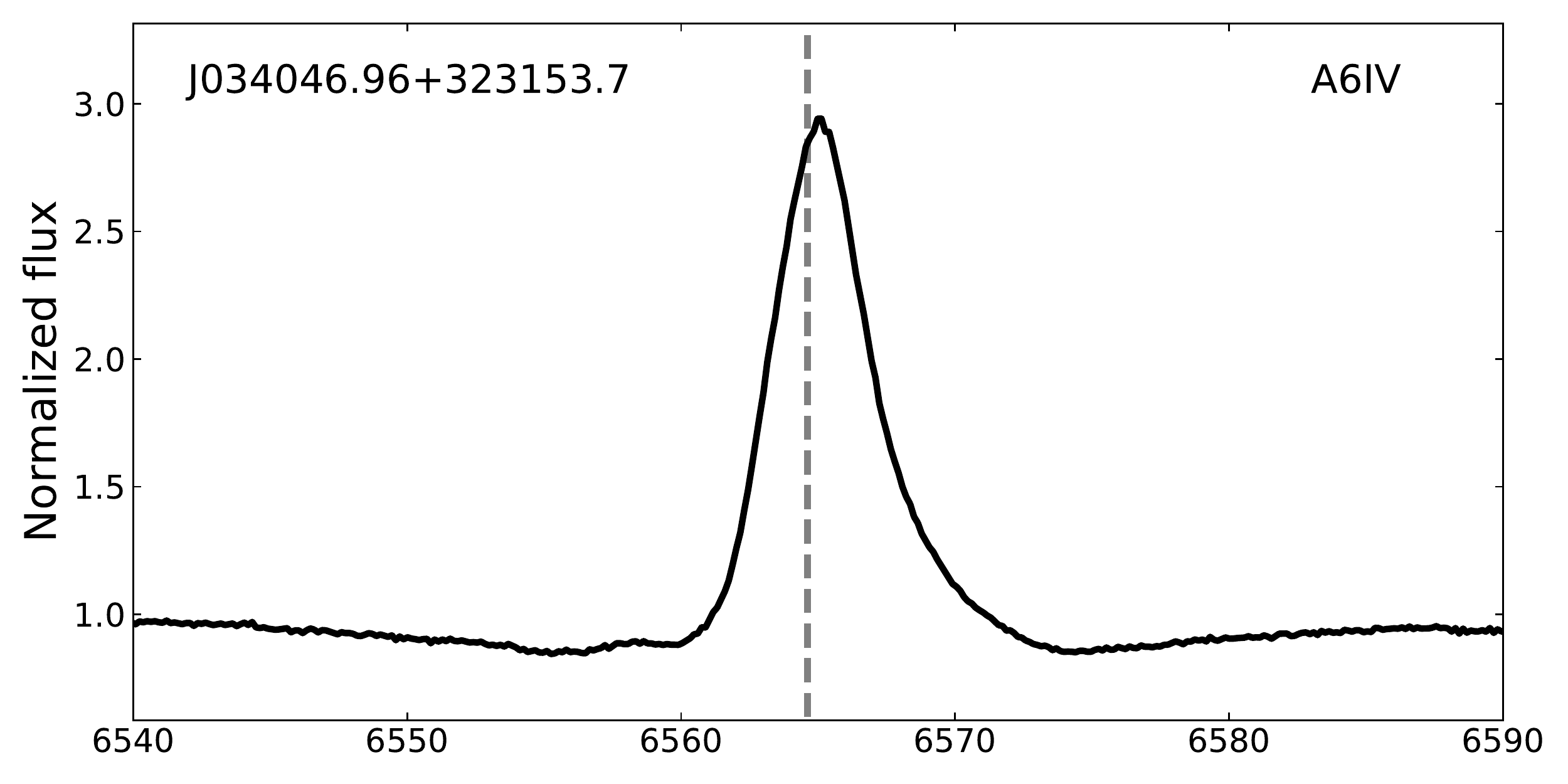}{0.5\textwidth}{J034046.96+323153.7}
          }
\caption{Two medium-resolution spectra with single exposure.
\label{fig:fig8}}
\end{figure}

We employed 66 confirmed HAeBes with medium-resolution spectroscopic survey from LAMOST DR7 and got 12 sources with available medium-resolution spectra, 3 of which are new HAeBes. Emission lines originated from star and nebulae can be distinguished well based on medium-resolution spectra such as H$\alpha$ emission of known HAeBe J064105.87+092255.6 shown in the top panel of Figure \ref{fig:fig8}. The H$\alpha$ emission line from the star is relatively wider while the H$\alpha$ emission of nebulae is much narrower as well as N\Rmnum{2}, and S\Rmnum{2} emission lines.

The object J034046.96+323153.7 has been observed in the medium-resolution spectra as shown in the bottom panel of Figure \ref{fig:fig8}. The counterpart of the low-resolution spectrum of this object has been masked within the region of H$\alpha$.

\subsection{HAeBes with P Cygni or inverse P Cygni profiles}
Seven of 66 confirmed HAeBes (6 new and 1 known) have P Cygni or inverse P Cygni profiles. Among them, J055054.77+201447.6,J035859.44+561112.6, J021547.65+441943.3, J060926.79+245519.8 are characterized by P Cygni profiles while J043152.11+475043.5, J213231.99+385754.9, J053135.15+095155.4 are characterized by inverse P Cygni profiles. According to \citet{Book.2007ASSL..342.....K}, the ratio of stars with such profiles is less than 20\% in HAeBes, and P Cygni profile appears more often than the inverse P Cygni profile, showing that the wind prevails over accretion. Our result agrees with that of \citet{Book.2007ASSL..342.....K}. They also pointed out stars sometimes show a change of state from wind to accretion or vice versa. \citet{PCchange.1993A&A...274..381P} reported the Herbig Ae star HR5999 showed a change of the profile of P Cygni into inverse P Cygni, suggesting the change from wind to accretion in envelope structure. To study the possible variation, follow-up time domain observations are needed.

\section{Discussion} \label{sec:discussion}

\subsection{HR Diagram \label{sec:HR}}


\begin{figure*}[ht!]
\begin{center}
\includegraphics[width=\textwidth]{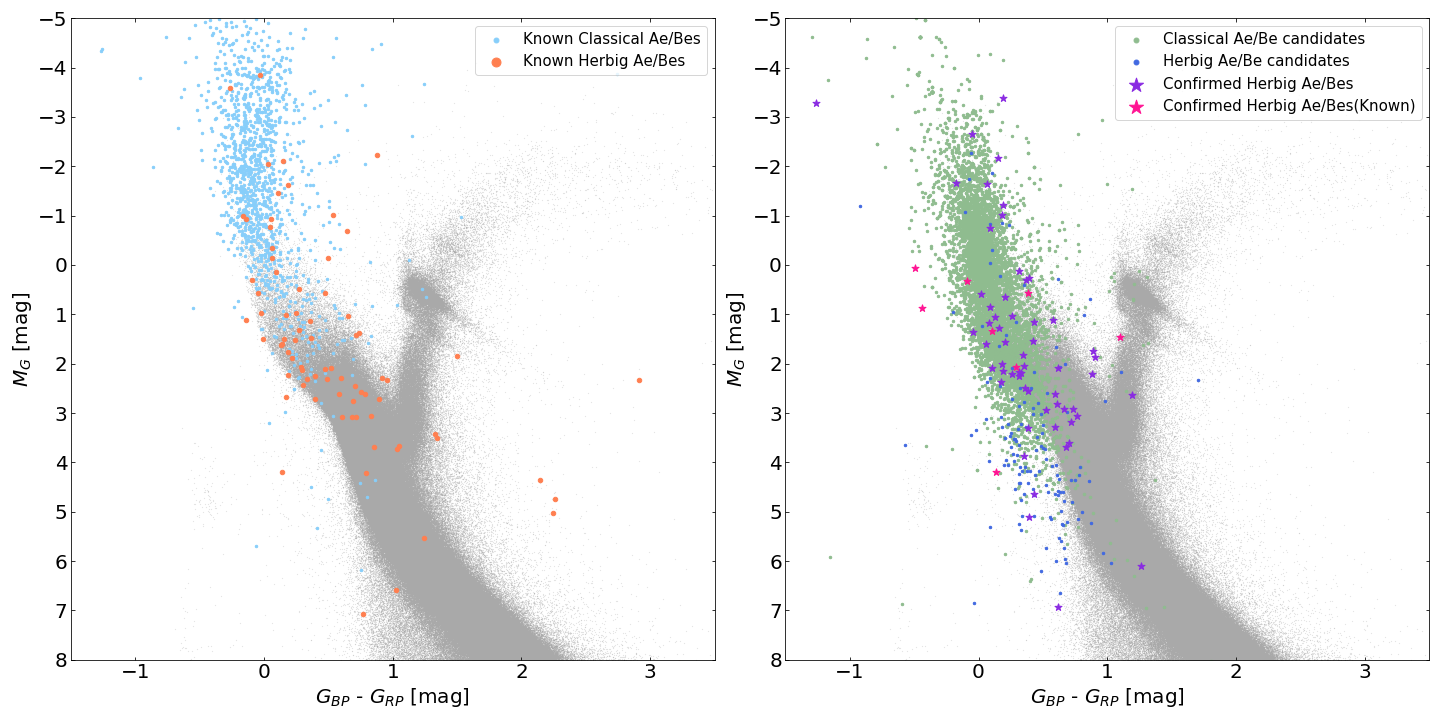}
\end{center}
\caption{Gaia color vs. absolute magnitude diagram. The left panel is the distribution of known HAeBes with good astrometric solution (($E(B-V)<0.2$), orange solid circle) and CAeBes (light blue solid circle). The right panel is the distribution of our confirmed HAeBes (pink star for known and violet star for new), HAeBe candidates (blue solid circle), and CAeBe candidates (green solid circle). The gray dots of two diagrams are 5 million sources of Gaia within 500pc distance. All colors have been corrected by interstellar extinction with 3d dust map.
\label{fig:fig9}}
\end{figure*}

Using estimated distances from \citet{Distance.2018AJ....156...58B} and Bayestar15 (\citet{Bayestar15.2015ApJ...810...25G}) as same as Subsection \ref{subsec:interstellar}, we derived E(B-V) of known HAeBes and CAeBes. After correcting the interstellar extinction using the extinction coefficients of Gaia (\citet{Extinction_gaia.2018MNRAS.479L.102C}), we calculated  the absolute magnitudes $M_G$ and plotted  known HAeBes on the HR diagram with interstellar extinction ($E(B-V)<0.2$). The HR diagram of known HAeBes and CAeBes is shown on the left panel of Figure \ref{fig:fig9}. The light blue solid circle represents a CAeBe and the orange one represents a HAeBe. The gray dots of the background are 5 million sources of Gaia within 500pc distance. Due to the dust envelopes/disks surrounding central stars, the absolute magnitude of HAeBe tends to be larger than CAeBe with similar temperatures. Also, there is a tendency that known HAeBes are more concentrated on the lower edge of the main sequence in the HR diagram. The HR diagram of our confirmed HAeBes and candidates  and CAeBes is shown on the right panel of Figure \ref{fig:fig9}. A green solid circle represents a CAeBe candidate and a blue one represents a HAeBe candidate. The star marks are confirmed HAeBes, pink of which are known while violet of which are newly identified. By comparing the two HR diagrams, it turns out HAeBes confirmed in our work have a similar distribution with the known ones. Besides, our HAeBe candidates are more concentrated at the lower edge of the main sequence. We will make further identification of the candidates in future work.

\subsection{Color distribution\label{sec:color}}

\begin{figure}[ht!]
\plotone{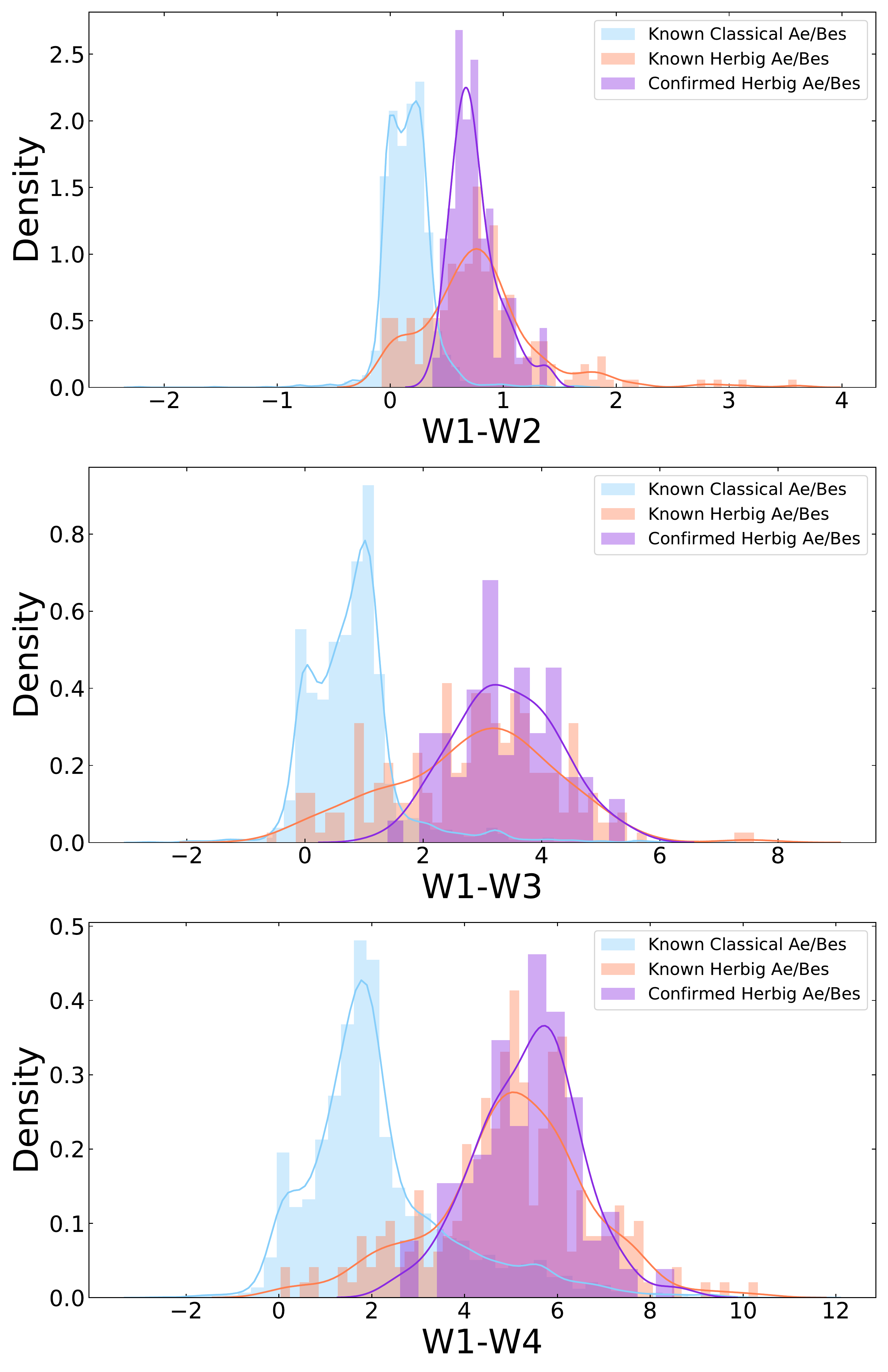}
\caption{WISE color distributions of HAeBes and CAeBes. 1,915 known HAeBes are marked with violet while 237 known CAeBes are marked with light blue. Violet color represents 66 confirmed HAeBes. The density and fitted curves are the results of Seaborn with a kernel density estimate (kde) in python.
\label{fig:fig10}}
\end{figure}

In order to estimate the effects of infrared excess for CAeBes and HAeBes, we chose colors of W1-W2, W1-W3, and W1-W4 as probes. The data we used includes 1,915 known CAeBes, 237 known HAeBes, and all of 66 confirmed HAeBes with complete data of the four photometric magnitudes. The color distributions of the three data sets are shown in Figure \ref{fig:fig10}. The color distributions of HAeBes confirmed by us are consistent with those of the known ones while quite different from that of known CAeBes. HAeBes tend to have much stronger excesses in mid-infrared than CAeBes.
Figure \ref{fig:fig10} suggests that W1-W4 color may be an additional criterion distinguishing HAeBes from CAeBes.

\subsection{Galactic distribution \label{sec:galactic}}

\begin{figure*}
\begin{center}
\includegraphics[width=\textwidth]{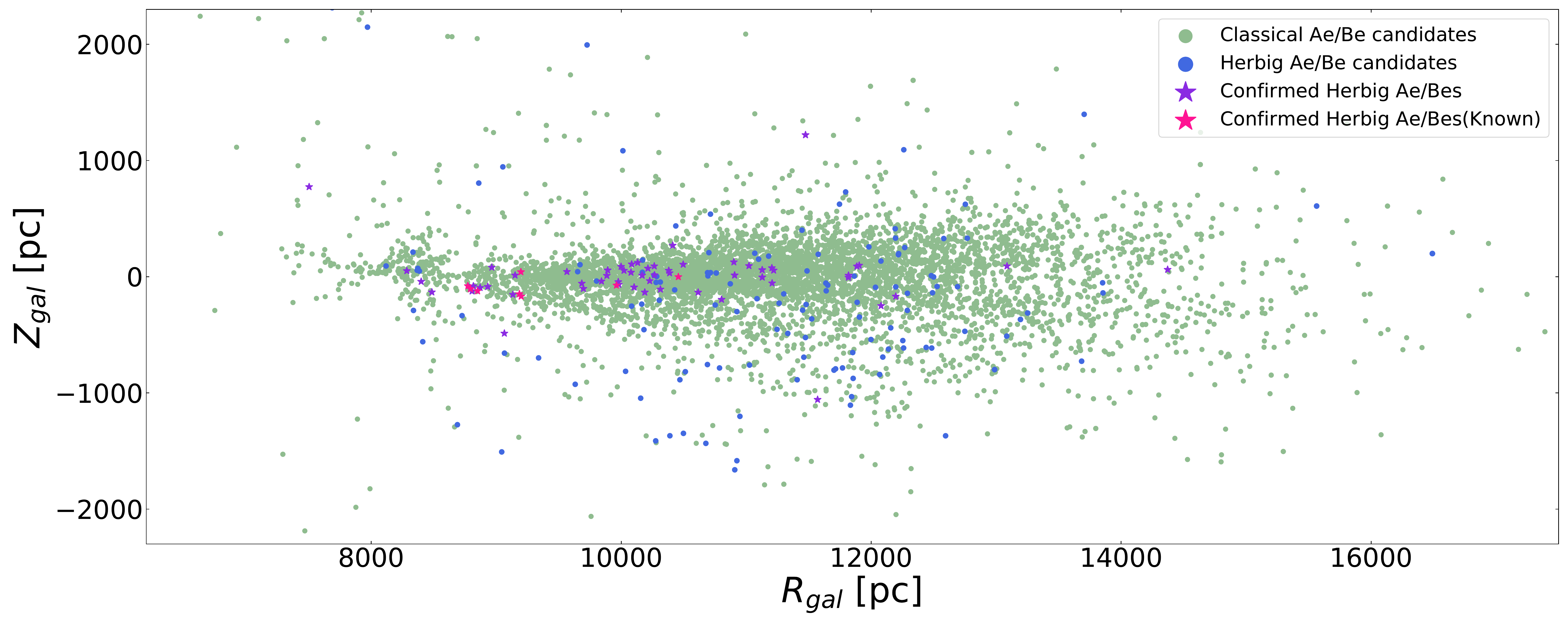}
\end{center}
\caption{Galactic distributions of confirmed HAeBe, HAeBe candidates, and CAeBe candidates. The color labels are the same as Figure \ref{fig:fig9}. The horizontal axis represents the projection distance $\rm \sqrt{X_{gal}^2+Y_{gal}^2}$ on Galactic disk to the center. The vertical axis represents the distance $\rm Z_{gal}$ to the Galactic disk.
\label{fig:fig11}}
\end{figure*}

Figure \ref{fig:fig11} shows the distributions of our confirmed HAeBes, HAeBe candidates, and CAeBe candidates in the Galaxy. The color labels are the same as Figure \ref{fig:fig9}. Most of our sources are located in the anti-center direction of Galactic due to the selection effect of observation. From the distance distribution, we can see that our stars are mainly located in the Galactic thin disk, and all HAeBes confirmed by us are located in the Galactic thin disk as well.

\begin{figure}
\plotone{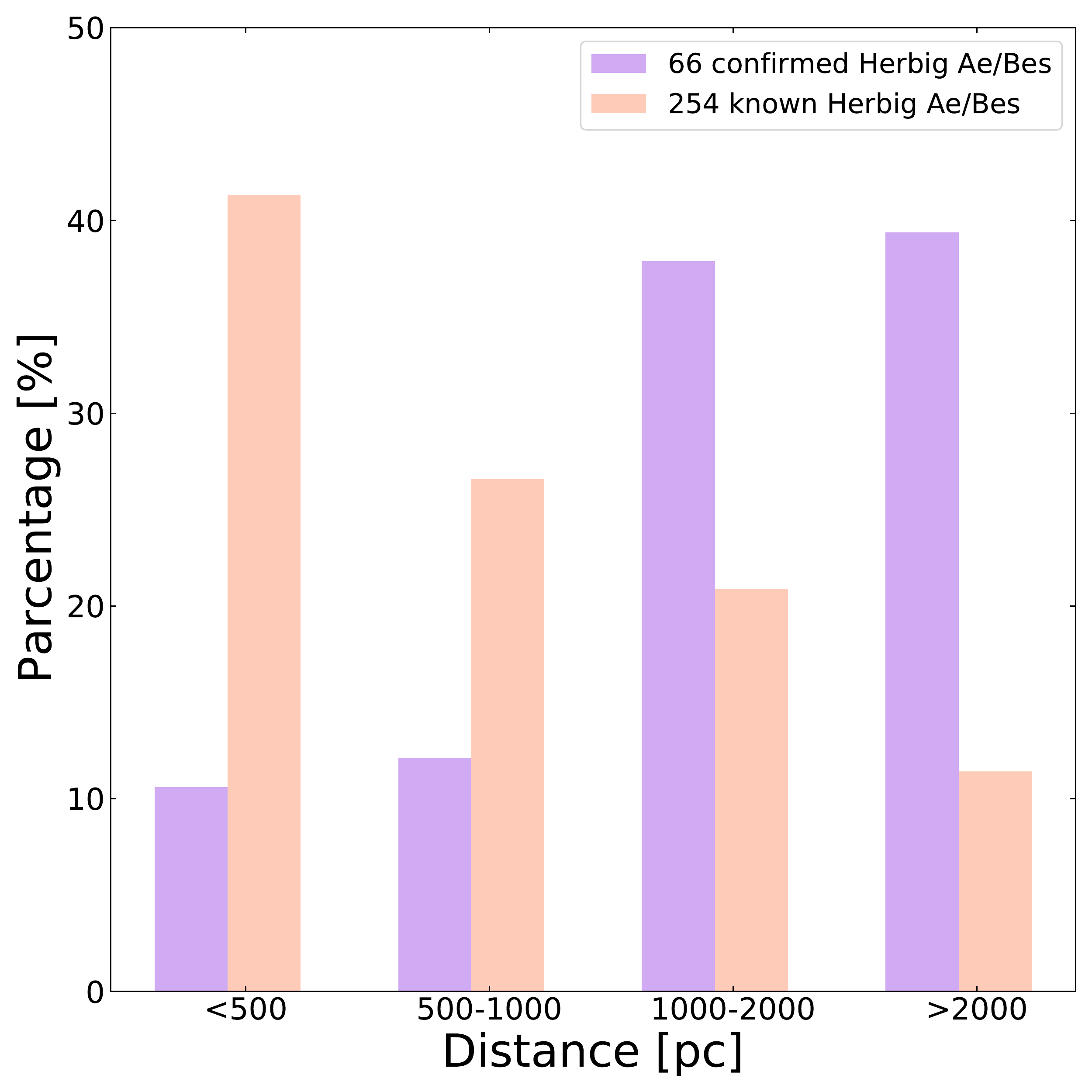}
\caption{Distances of known and this work confirmed HAeBes to the Sun. The vertical axis represents the percentage of each set.
\label{fig:fig12}}
\end{figure}

Figure \ref{fig:fig12} shows the distances of known and confirmed HAeBes to the Sun. 77.3\% of the conformed HAeBes are farther than 1Kpc, while 67.7\% of the known HAeBes are within 1Kpc. It means our HAeBe sample enlarges the number of known HAeBe in further distance.

\subsection{HAeBe candidates with nebula emission  \label{subsec:nabulae HAeBes}}

Apart from the 66 confirmed HAeBes, we also have a fraction of HAeBe candidates with nebula emission  lines of N\Rmnum{2} and O\Rmnum{3}. To ensure stellar intrinsic H$\alpha$  emission lines showing in spectra, medium or high-resolution spectra are necessary because the wide stellar H$\alpha$  emission lines and narrow nebulae emission lines can be clearly set apart in medium or high-resolution spectra. A planetary nebula (PN) as the main interference shows narrow emission lines in spectra and IR excess of W4 magnitudes. The core of a PN is usually a white dwarf with an expanding, glowing shell of ionized gas. As a crucial definition of HAeBe, the confirmation of star origin H$\alpha$ emission is required here. Except for 66 confirmed HAeBes, some of our candidates might be PNe, for example, the targets J070922.52-004823.6 and J072001.54+181725.9 have been identified as PNe in the Simbad database.  On the contrary, targets classified as YSO candidates and emission-line stars in Simbad are probably true HAeBe, such as J052616.37+345333.1 with an ID of 2MASS J05261637+3453332.


\section{Summary \label{sec:summary}}
In this paper, we provide a catalog of 30,048 spectra for early-type emission-line stars from LAMOST DR7. We divided emission-line spectra into 10 subclasses of 3 main classes based on the morphologies of H$\alpha$ emission-line profiles and flagged the spectra contaminated by nebula emission  lines. Furthermore, we estimated velocities of stellar winds or accretion flows for 103 stars with P Cygni or inverse P Cygni profiles in the spectra. Moreover, we proposed a new method to separate HAeBes from CAeBes based on two color-color diagrams (H-K,J-H) and (H-K,K-W1) and applied the new method to our catalog. In total, we got 235 spectra of 201 HAeBe candidates along with 6,741 spectra of 5,547 CAeBe candidates. We also confirmed 66 HAeBes from 201 candidates based on spectra of LAMOST DR7 and images of WISE, 58 of which are newly identified.

There are 4 catalogs provided in this paper, including a catalog of H$\alpha$ emission-line stars (Table \ref{tab:general}), a catalog of stellar wind and accretion flow velocities (Table \ref{tab:velocities}), a catalog of 66 confirmed HAeBes (Table \ref{tab:confirmed}), and a catalog of HAeBes associated with solar-nearby SFRs (Table \ref{tab:nebulae}). All the full catalog would be provided by the corresponding Author.

In the general catalog, 22,238 spectra of 18,965 emission-line stars and 7,810 spectra of 6,921 emission-line star candidates are provided by us as shown in Table \ref{tab:general}. The 'SpecID' column is the unique ID of spectra.  The 'Designation' column represents LAMOST source ID. The 'Simbad\_ID' column represents ID from the Simbad\footnote{\url{http://simbad.u-strasbg.fr/simbad/}}. The column 'LAMOST\_class'  and 'rv' are from LAMOST 1D-pipeline result. The column 'H$\alpha$ type' represents the type of H$\alpha$ emission-line profile based on morphological classification. The column 'Objtype' is the result of our work, including classical Ae/Be candidates, HAeBes, HAeBe candidates, EM and EM candidate. The column 'Objtype\_literature' is the result in literatures including 4 classes (CV, Algol-type binary, CAeBe, and HAeBe). The column 'IF\_n' is the result of our work to show if nebula emission  line exist. A 'Y' marked in the '$\rm IF\_n$' column means nebula emission  lines are identified in the spectrum. A 'Y' marked in the '$\rm IF\_H\Rmnum{2}$' column means the star locates in the known H\Rmnum{2} regions provided in the catalog above.

\begin{splitdeluxetable*}{cccccBccccccc}
\tablenum{8}
\tablecaption{Catalog of H$\alpha$ emission-line stars \label{tab:general}}
\tablewidth{0pt}
\tabletypesize{\scriptsize}
\tablehead{
\colhead{ } & \colhead{ } & \colhead{ } & \colhead{ra} & \colhead{dec} & \colhead{ } & \colhead{rv} & \colhead{} & \colhead{ } & \colhead{ } & \colhead{ } & \colhead{ }\\
\colhead{Specid} &\colhead{Designation} & \colhead{SIMBAD\_ID} & \colhead{[deg]} & \colhead{[deg]} & \colhead{LAMOST\_class} & \colhead{[km/s]} & \colhead{H$\alpha$ type} & \colhead{Obj\_type} & \colhead{Objtype\_literature} & \colhead{$\rm IF\_n$} & \colhead{$\rm IF\_H\Rmnum{2}$}
}
\startdata
20141120HD000408N565515B0104170 & J000109.17+565053.6 &     & 0.2882314 & 56.8482265 & A7V & -79.2411425 & 1.3 & EM &   & Y &\\
20141120HD000408N565515B0104198 & J000143.10+563555.4 &     & 0.4296194 & 56.5987436 & A1IV & -71.68337463 & 1.3 & classical Ae/Be candidate &  &   &   \\
20141218HD000308N424452V0105040 & J000228.06+414202.5 & IRAS 23599+4125 & 0.616952 & 41.700721 & A7V & -76.12989637 & 2.3 & EM &  &   &   \\
20141120HD000408N565515V0111195 & J000300.22+585223.9 & TYC 3664-695-1 & 0.750944 & 58.873321 & A2V & -35.21062419 & 2.3 & classical Ae/Be candidate &  &   &   \\
20141120HD000408N565515V0115104 & J000352.84+573550.0 &     & 0.9702061 & 57.5972327 & B6 & -53.08724846 & 1.2 & classical Ae/Be candidate &  &   &   \\
20171016GACII000N48B111248 & J000419.24+500327.2 &   & 1.0801791 & 50.057578 & A7IV & 40.38504202 & 1.2 & EM &  & Y &  \\
20131030VB002N48V103056 & J000507.49+482705.0 & UCAC4 693-000644 & 1.28123 & 48.451401 & A3IV & -12.21654266 & 2.1 & classical Ae/Be candidate  &   &  & \\
20131122HD235547N481312V0113213 & J000507.49+482705.0 & UCAC4 693-000644 & 1.28122 & 48.451398 & A3IV & -11.99169832 & 2.1 & classical Ae/Be candidate  &   &  & \\
\enddata
\tablecomments{The 'SpecID' column is the unique ID of spectra. The column 'Objtype' is the result of our work, including classical Ae/Be candidates, HAeBes, HAeBe candidates, EM and EM candidate. The column 'Objtype\_literature' is the result in literatures including 4 classes (CV, Algol-type binary, CAeBe, and HAeBe). A 'Y' marked in the '$\rm IF\_n$' column means nebula emission  lines are identified in the spectrum. A 'Y' marked in the '$\rm IF\_H\Rmnum{2}$' column means the star locates in the known H\Rmnum{2} regions provided by \citet{WISE.2014ApJS..212....1A}. Complete list is available in electronic form.}
\end{splitdeluxetable*}


\section*{Acknowledgements}

This work is supported by National Science Foundation of China (Nos U1931209, 12003050) and National Key R\&D Program of China(No. 2019YFA0405502). Guoshoujing Telescope (the Large Sky Area Multi-Object Fiber Spectroscopic Telescope, LAMOST) is a National Major Scientific Project built by the Chinese Academy of Sciences. Funding for the project has been provided by the National Development and Reform Commission. LAMOST is operated and managed by the National Astronomical Observatories, Chinese Academy of Sciences.

\bibliography{ref}{}
\bibliographystyle{aasjournal}



\end{document}